\documentclass[11pt]{article}
\usepackage[utf8]{inputenc}
\usepackage[T1]{fontenc}
\usepackage[margin=2.2cm]{geometry}
\usepackage{amsmath,amssymb}
\usepackage{graphicx}
\usepackage{xcolor}
\usepackage{framed}
\usepackage{enumitem}
\usepackage{caption,subcaption}
\usepackage[colorlinks=true,linkcolor=blue,urlcolor=blue,citecolor=blue]{hyperref}

\setlist{nosep,leftmargin=1.4em}
\newcommand{\post}[1]{\texttt{#1}}
\newcommand{\fbar}{\bar f}
\newcommand{\Feff}{F_{\mathrm{eff}}}
\definecolor{mscolor}{RGB}{0,160,220}%

\newif\ifinternalpars\internalparstrue%
\newcommand{\ipar}[1]{\ifinternalpars\paragraph{#1}\fi}

\usepackage{authblk}

\title{Predictability of Collective Behavior in Networks of {LLM} Agents}

\author[1]{Fabian Baumann\thanks{baumannf@sas.upenn.edu}}
\author[2]{Michele Starnini\thanks{michele.starnini@gmail.com}}

\affil[1]{University of Pennsylvania, Philadelphia, PA 19104, USA}
\affil[2]{Department of Engineering, Universitat Pompeu Fabra, 08018 Barcelona, Spain}

\date{}

\begin{document}
\maketitle
\begin{abstract}
{As AI agents built on large language models (LLMs) increasingly interact at scale, anticipating their collective behavior becomes necessary. Unlike humans, individual LLM agents can be probed in {controlled scenarios}; however, the predictability of their collective outcomes hinges on what such measurements capture. We study information cascades on networks of LLM agents that decide whether to share a post given how many of their contacts have shared it. At the same fraction of sharing contacts, an agent's response depends on how many contacts it has and on the post. Responses that ignore the number of contacts mispredict the cascades, increasingly on heterogeneous networks, by an error we derive analytically and that network topology alone cannot reveal. Measured in context, the response predicts, through mean-field theory, the cascades of full LLM simulations, tipping points included. Anticipating the collective behavior of AI agents thus requires evaluating them in their local network context.}
\end{abstract}

\section{Introduction}
\label{sec:motivation}
\label{sec:scope}

Large numbers of AI agents have started to act autonomously on our behalf, including shopping, searching for
information, and posting on social media \cite{yang2025agenticwebweavingweb,li2024personal}. LLM-driven AI agents interact not only with humans but with one another, a phenomenon one could recently witness in a number of cases ranging from social media platforms populated by AI agents \cite{price2026claws,li2026rise} to cybersecurity incidents caused by swarms of agentic AI systems \cite{greenblatt2026investigation}.
It is natural to ask whether such
populations reproduce the collective phenomena that human populations do \cite{brinkmann2023machine}{, from the emergence of shared
conventions and consensus \cite{castellano2009statistical,centola2015spontaneous,ashery2025emergent} to polarization \cite{baumann2020modeling,piao2025emergence} and the viral spreading of information \cite{vosoughi2018spread,yang2024oasis,liu2025mosaic,zhang2025llmaidsim},} and, crucially, under
what circumstances and how precisely {their collective} behavior can be predicted. The question is of practical importance: when agents share content on our behalf, whether a post spreads is decided by
the collective dynamics of the population, not by the choice of any single agent, and such an outcome should be
anticipated before it happens rather than observed after the fact \cite{hammond2025multi,schroeder2026malicious,flint2026indirect}.

Predicting the behavior of a collective from that of its members is the classic micro-to-macro problem of the
social sciences and of statistical physics \cite{schelling1978micromotives,granovetter1978threshold,castellano2009statistical,starnini2025opinion}. In human populations it is often hard to bridge this micro-macro gap, because the response of an individual to its social environment can rarely be measured directly{: outside} small controlled experiments \cite{centola2015spontaneous,centola2018experimental,andreoni2021predicting}{, it} must be inferred from observational data~\cite{romero2011differences}. This obstacle does not exist for LLM agents.  The response of a single agent to its (social) environment can be measured directly, by describing that environment to the agent in a prompt and querying the LLM as often as needed to estimate the probability distribution of answers. Each ingredient of the environment, from the number of contacts to the content the agent is shown, can then be varied on its own while all the others are held fixed, as in a controlled experiment, but with a precision and at a scale that experiments with humans cannot reach. Moreover, an LLM agent given the same prompt behaves the same in a measurement as in a population, whereas human behavior in the laboratory can differ from that outside it~\cite{levitt2007laboratory}.
A promising wave of recent work has used this methodology to demonstrate that the collective behavior of LLM populations can indeed be predicted from the individual agent's behavior{, building on the observation that LLM populations spontaneously develop shared conventions \cite{ashery2025emergent}}.
{The idea is to probe how a single agent responds to different environmental cues, and use mean-field theory to recover the macroscopic outcome, as demonstrated for majority-following and spontaneous coordination \cite{de2026ai}, for group-size effects in naming games \cite{flint2026group}, and for opinion dynamics with a fitted Ising-like response \cite{el2026physics}.} 

Most of these works, however, share a structural simplification. Agents are embedded in a well-mixed population, in which every agent can interact with every other, so that each agent responds, directly or through randomly drawn partners, to the state of the population as a whole, such as the average opinion \cite{de2026ai} or the overall fraction of agents that have adopted a norm \cite{flint2026group}. Recent work places the responses measured in this setting on networks \cite{passaro2026network}, assuming that they do not depend on the agent's position in the network. In realistic settings, by contrast, LLM agents will often have access to information about their immediate environment, such as what the accounts they follow have posted or shared, and this local information may shape their behavior more than any global signal. If it does, the network in which the agents are embedded becomes decisive for the collective dynamics, because it determines the information each agent is exposed to.
In many collective phenomena, from the adoption of a convention to the sharing of a post, an agent's decision {may then depend} on how many of its own contacts have already adopted a behavior. Since agents on a network differ in how many contacts they have, it matters whether an agent responds to the fraction or to the number of these contacts. Threshold models of contagion have long settled this question by assumption \cite{granovetter1978threshold,watts2002simple,dodds2004universal,centola2007complex}; for LLM agents it can be measured directly. {Experiments in well-mixed groups already show that how strongly LLM agents follow a given majority decreases as the group grows \cite{de2026ai}.}

This raises a general question about populations of LLM agents. \emph{How is the behavior of a single agent linked to that of the population, and which aspects of a single agent's response must be measured to predict the collective behavior?}
In principle, the response of an LLM agent can depend on everything in its prompt, from how many of its contacts have shared and how many contacts it has to the content of the post, its memory of past interactions \cite{park2023generative,dodds2004universal,ashery2025emergent}, the persona or ideology it is given \cite{argyle2023out,piao2025emergence} and the incentives of the agent, such as reward maximization \cite{akata2025playing}, as well as on the language model \cite{nohr2026peer} and its biases \cite{santurkar2023whose,chuang2024simulating}. Each variable left out of the measurement is a potential source of error in the prediction. We address the question for information cascades, for which the theory linking individual responses to collective outcomes on networks is mature \cite{watts2002simple,dodds2004universal,gleeson2007seed,gleeson2013binary}, so that comparing the prediction with full LLM simulations tests the measured response rather than the theory. We measure the response as a function of the most structural of these variables, how many contacts an agent has and how many of them have shared, for different posts, and keep the others fixed.

{We show that the collective behavior of interacting LLM agents can be predicted from measurements of a single agent, provided the measurement captures the agent's local social context. We study information cascades on social networks of LLM agents, in which each agent decides whether to share a post after seeing how many of the accounts it follows have shared it.} 
First, we find that networked LLM populations produce information cascades whose reach depends on both the network and the post. The decision of each agent depends on its local context in a way that no single response curve can describe. Agents who follow more accounts share when a smaller fraction of them has shared: an agent that sees 2 of its 10 contacts share a post almost never shares it, while one that sees 20 of its 100 contacts share it does so far more often. The response also changes with the content of the post. Second, we show that responses that ignore the number of contacts mispredict cascades, increasingly so on more heterogeneous networks and in a direction that depends on the content, whether they depend on the fraction of sharing contacts alone, as threshold models assume, or are measured in a single large group, as in previous studies. Third, the response measured in context, fed into the mean-field theory of cascades on networks, predicts the cascades of full LLM simulations, including {whether they spread through the network and the seed fraction at which they tip}; once the response is measured, the cascade on any network follows without further queries to the LLM. This agreement is expected for agents that decide only on what they are shown, provided the measurement records what they respond to; the failure of the responses that ignore the number of contacts shows that this condition is not met by default. Finally, we derive the error of these responses analytically, for any response that depends on the number of contacts, and show that it cannot be inferred from network statistics alone. The collective behavior of LLM populations is thus predictable, but only if each agent is measured in the
social context in which it acts.

\section{Results}
\subsection{\texorpdfstring{{Information cascades in networked LLM populations}}{Information cascades in networked LLM populations}}
\label{sec:model}

\begin{figure}[tbp]
  \centering
  \includegraphics[width=0.95\textwidth]{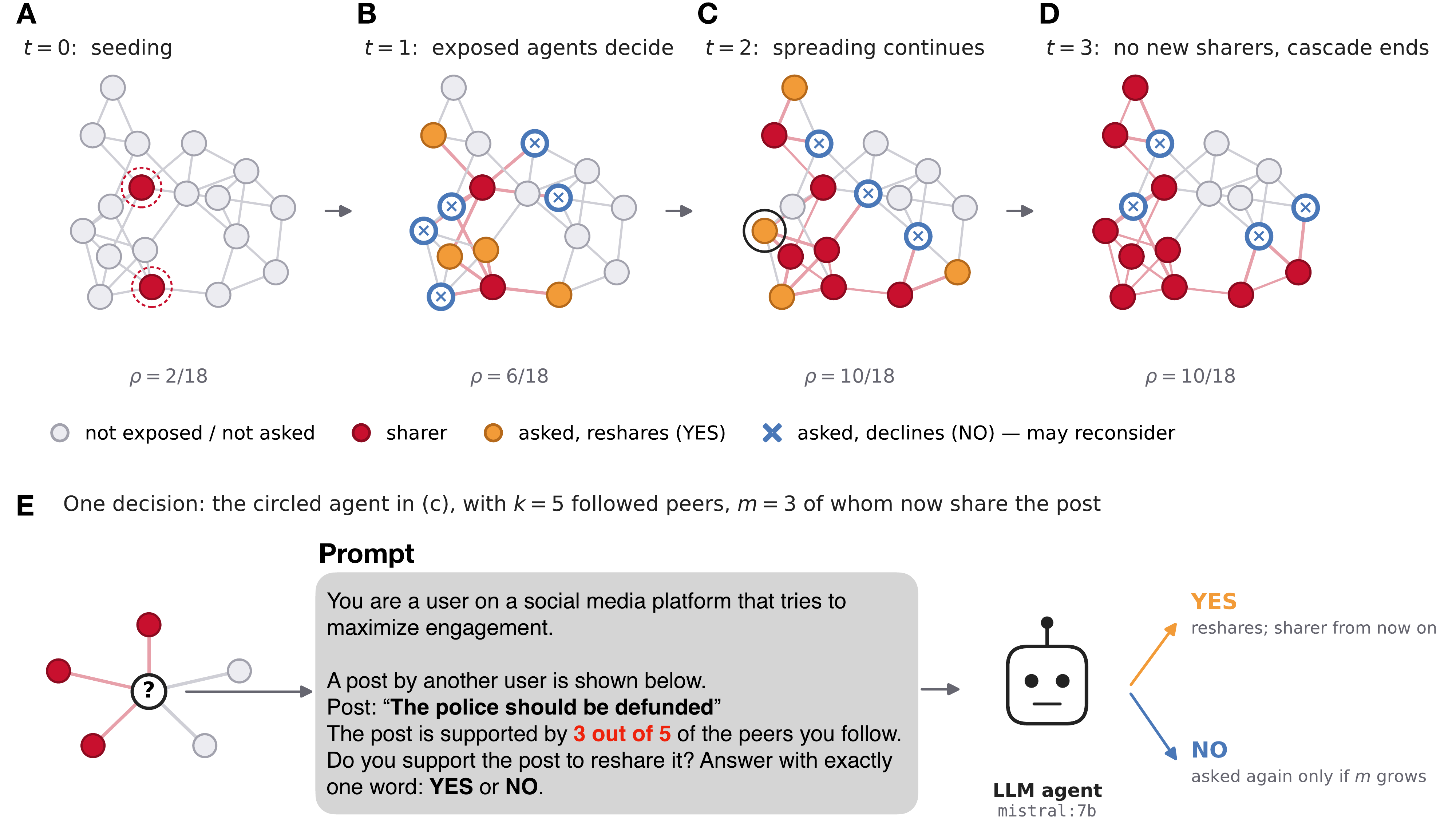}
  \caption{\textbf{Information cascades in a population of LLM agents.} (\textbf{A} to \textbf{D}) Spreading from the seed agents ($t=0$) to the end of the cascade ($t=3$), when no agent shares any more. (\textbf{E}) A single decision: the prompt tells the agent how many of the accounts it follows have shared the post, and the agent shares it (YES) or not (NO); an agent that declines is asked again only if more of its contacts share.}
  \label{fig:schematic}
\end{figure}

\begin{figure}[tbp]
  \centering
  \includegraphics[width=0.99\textwidth]{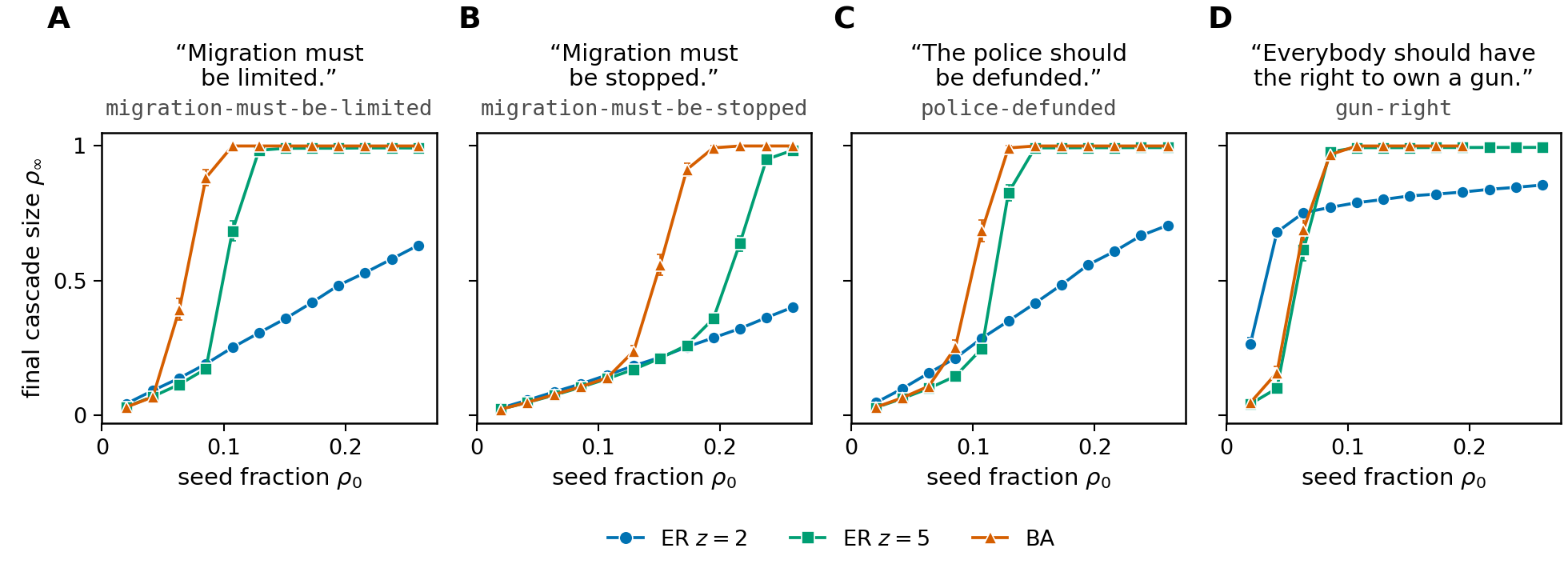}
  \caption{\textbf{Information cascades in LLM populations depend on both the network and the post.} Final cascade size $\rho_\infty$ against the seed fraction $\rho_0$ in the LLM population, an LLM agent at every node, on Erd\H{o}s--R\'enyi (ER) networks with mean degree $z=2$ and $z=5$ and on Barab\'asi--Albert (BA) networks, all with $N=1000$ agents. (\textbf{A} to \textbf{D}) The four posts; each panel gives the text of the post and the name used in the text. For three posts the cascade spreads most easily on BA and least on the sparse ER network; for \post{gun-right} the sparse ER network is the one on which the smallest seed already triggers a large cascade. Each point averages 100 runs; the error bars (standard errors) are mostly smaller than the symbols.}
  \label{fig:llmresults}
\end{figure}

\ipar{Exposure-triggered sharing dynamics}\label{sec:setup}
We simulate the spreading of four different social media posts on a social network of {LLM} agents; the process is schematically depicted in Fig.~\ref{fig:schematic}. Specifically, we study four posts on contested political topics: two posts about
migration, one about defunding the police and one about gun rights (Fig.~\ref{fig:llmresults}). 
{We chose them because the spread of such posts on social media matters}~\cite{brady2017emotion}. 

The LLM that governs the decisions of our agents is the open-weight model \texttt{mistral:7b}; it gives us
full control over the prompt and the decoding, and makes every measurement reproducible.

We consider a network of $N$ LLM agents, {one at each node of the social network,} where each agent follows its $k$ neighbors.
At the start of the dynamics, a fraction $\rho_0$ of agents, chosen uniformly at random, share the post (seed agents). 
The cascade then unfolds in discrete rounds. An agent decides whether to {share} the post whenever something changes in its local environment, i.e., when the number of its neighbors sharing the post has increased since it last made a sharing decision. 
Agents with no sharing neighbors never encounter the post, and therefore never get to make a sharing decision. 
An agent that declines is not excluded from the process: It may reconsider once further neighbors share the post, so repeated exposure gives a post several opportunities to be adopted. 
This mirrors how content reaches users on social media, where a post {can reach a user again} each time another account they follow shares it.
All decisions within a round are made simultaneously, based on the state of the network at the beginning of that round. 
Newly sharing agents then raise the exposure of their own neighbors, who are asked in the following round (see Fig.~\ref{fig:schematic}). 
The cascade ends when a round produces no new sharers. We record the final fraction of sharers, $\rho_\infty$, as the size of the cascade.

Each decision of an LLM agent is a single, independent query to the language model {(Materials and Methods)}. The model is instructed to act as a user on a social media platform that tries to maximize engagement. It is shown the post and told that the post is supported by $m$ out of the $k$ peers it follows, and it is asked to answer YES or NO to whether it {shares}; the prompt is depicted in Fig.~\ref{fig:schematic}E. Queries carry no memory: an agent receives no information about its own earlier decisions or about the history of the cascade beyond the current count $m$. Replies are parsed for YES or NO{, and a reply that contains neither is handled as described in Materials and Methods}.

\ipar{Interplay of network topology and post content}\label{sec:observed}

The results of the full LLM simulations are shown in Fig.~\ref{fig:llmresults}. {Each network has $N=1000$ agents, and for each network and post we run 100 cascades at each of 12 seed fractions (Materials and Methods).} Both the final cascade size $\rho_\infty$
and the seed fraction $\rho_0$ required to trigger a global cascade depend strongly on the network structure and on the content of the post. 
{We call a cascade global when it reaches at least half of the agents{, and the tipping seed $\rho_0^{*}$ the seed fraction at which the cascade becomes global}.}  The effect of the network can be judged within each panel, where the different symbols (and colors) correspond to different network structures. Generally, global cascades are triggered more easily in Barab\'asi--Albert networks than in Erd\H{o}s--R\'enyi networks of comparable mean degree, and within Erd\H{o}s--R\'enyi networks they become easier to trigger as the mean degree increases{, except for \post{gun-right}, whose cascade reaches half of the agents from a smaller seed on ER $z=2$ ($\rho_0\approx0.03$) than on ER $z=5$ ($0.06$)}. The content of the post, depicted in panels A to D, also introduces considerable variation in the seed fraction needed to trigger {a} global cascade{: across posts it varies by a factor of 2.6 on BA (from $0.056$ to $0.147$) and 3.5 on ER $z=5$ (from $0.059$ to $0.206$)}. 

\subsection{\texorpdfstring{The response of an agent depends on its degree and the post content}{The response of an agent depends on its degree and the post content}}
\label{sec:kernel}
\label{sec:model-def}

\ipar{Measuring the response of a single agent}\label{sec:measure} 
To better understand, and ultimately predict, information cascades in realistic {LLM} populations, we construct reduced models of the system.  To this end, we first measure the sharing statistics of individual {LLM} agents in terms of a function $f(m,k)$, defined as 
\begin{equation}
  f(m,k)\;=\;P\!\left(\text{share at a single exposure}\;\middle|\; m \text{ of } k \text{ followed accounts have shared}\right);
\end{equation}
the accounts an agent follows are its neighbors on the network.

The idea is to replace {LLM} agents (making sharing decisions about posts) with simpler probabilistic agents that have {the same} information-sharing statistics.
Since each query contains only the post, $m$ and $k$, $f(m,k)$ measured for each post {determines the decisions of the agent} in our setting. The same measurement can be conditioned on any further variable an agent may respond to, such as its memory of past interactions, the persona or ideology it is given, or the model it runs; here we focus on the variables $m$ and $k$.
The function $f(m,k)$ is a probabilistic
(``soft'') response: it can take any value between 0 and 1, rather than jumping from 0 to 1 at a threshold. 
We measure $f(m,k)$ by presenting individual agents with exactly the prompts $P(\mathrm{post},m,k)$ used in the full population simulations (Fig.~\ref{fig:schematic}E). For each combination of post, $m$ and $k$, we query the LLM 200 times and estimate $f(m,k)$ as the fraction of YES responses, i.e., the probability that an agent {shares} the post in a single decision when $m$ of its $k$ followed peers have shared it.
The response is measured for 147 degrees between $k=1$
and $k=155$, which includes every degree that occurs in our networks (the largest is {$155$}, on BA), so it is never used beyond the degrees at which it was measured.

\begin{figure}[tbp]
  \centering
  \includegraphics[width=0.95\textwidth]{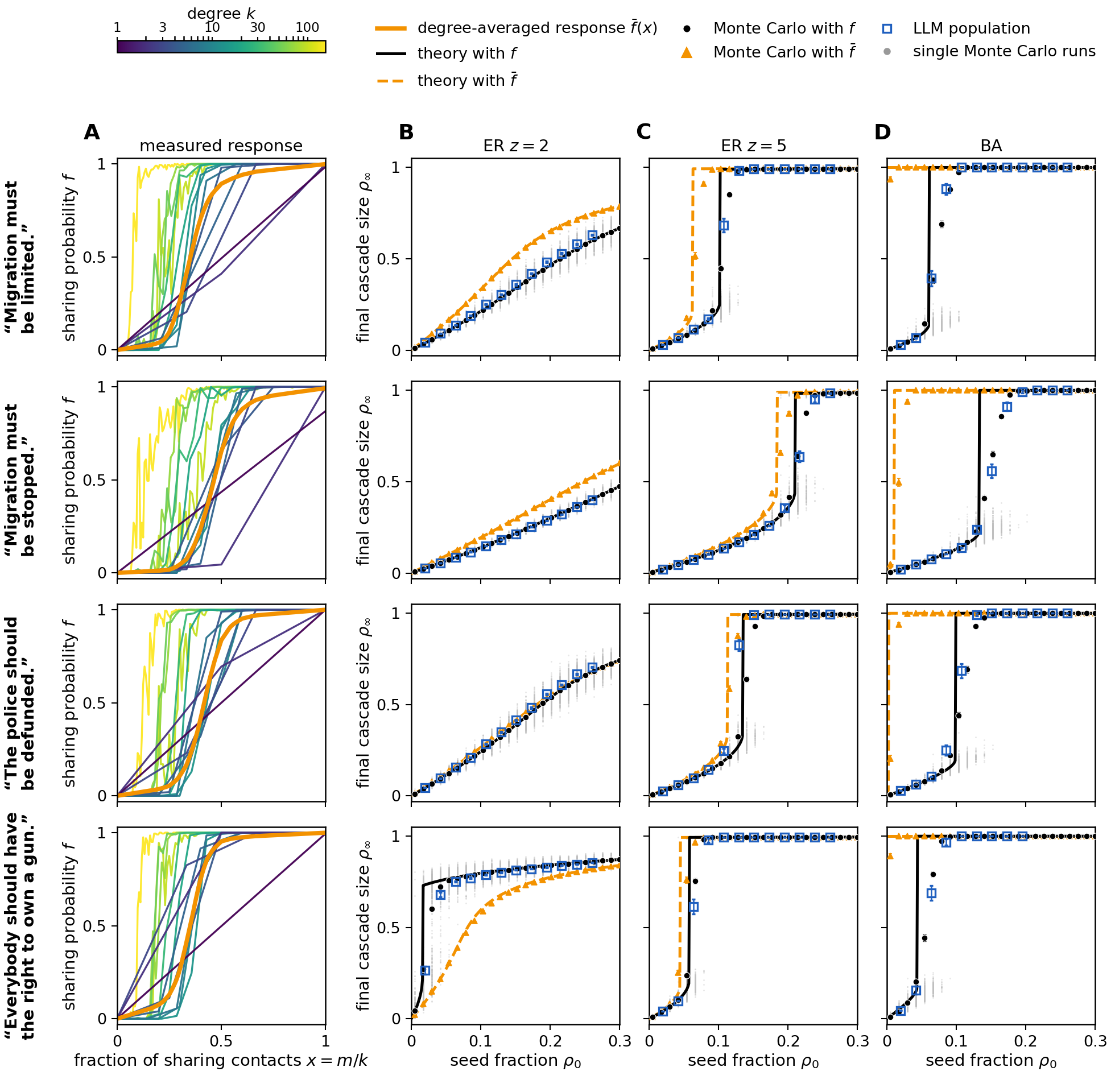}
  \caption{\textbf{The measured response predicts the cascades; its degree average does not.} Rows: the four posts, as in Fig.~\ref{fig:llmresults}. (\textbf{A}) Measured probability of sharing $f(m,k)$ against the fraction $x=m/k$ of sharing contacts, one curve per degree $k$, for 14 degrees between $k=1$ and $k=152$ spaced logarithmically (color scale at the top); orange: the degree-averaged response $\fbar(x)$ of the ER $z=5$ network. (\textbf{B} to \textbf{D}) Final cascade size $\rho_\infty$ against the seed fraction $\rho_0$ on ER $z=2$ (continuous transition), ER $z=5$ (discontinuous) and BA, all with $N=1000$. Lines: mean-field theory with the degree-resolved response $\Feff[f]$ (black) and with the degree-averaged response $\Feff[\fbar]$ (orange, dashed). Symbols: Monte Carlo averages with the same two responses (black circles, orange triangles), single Monte Carlo runs (gray dots) and the LLM population (blue squares; error bars: standard errors). The theory matches the Monte Carlo {simulations} for both responses on every network, and the LLM population follows the Monte Carlo {simulations} with $f$, except for a slightly earlier spread of \post{police-defunded} and, to a lesser extent, of \post{migration-must-be-limited}. With $\fbar$, the first three posts ignite at a lower seed on both ER networks, and from a very small seed on BA. For \post{gun-right} on ER $z=2$ the direction reverses: only the cascade with $f$ tips (Section~\ref{sec:content}).}
  \label{fig:cascade}
\end{figure}

\ipar{The response depends on degree}\label{sec:degree}
Fig.~\ref{fig:cascade}A shows the measured response functions $f(m,k)$ for each post and for several values of $k$. Our first and most important observation is that $f$ does not depend on the ratio $m/k$ alone: the curves for different $k$ do not collapse onto a single curve.\footnote{This holds for the measured points themselves, not only for the lines that join them: the line of each
degree only interpolates between its points $x=0,1/k,\dots,1$, so lines of different degrees would not coincide exactly
even if all the points fell on one curve; but the points do not fall on one curve either --- at the same fraction $x$,
agents of different degree share with different probabilities.} Instead, $f$ depends strongly on the degree $k$, and for higher-degree agents it typically saturates at a smaller fraction $m/k$ of sharing peers. Second, $f(m,k)$ increases monotonically with $m$ only for small degrees; for larger $k$ it becomes non-monotonic. 
Third, the response functions differ considerably between posts. For example, {at most degrees,} more social pressure (a larger $m$) is needed to raise the sharing probability of ``Migration must be stopped.'' than that of ``Migration must be limited.''

For all four posts, the probability of {sharing}
depends strongly on the social signal: it is close to zero when only a small fraction of the agent's contacts has
shared, and close to one when all have (Fig.~\ref{fig:cascade}A), so the agents do take into account
what their contacts did{, and the probability usually rises gradually rather than in a single step}.
The non-monotonicity at large $k$ is a small modulation of this overall increase with $m$; in any case, the analysis presented below does not require any particular shape of this response: the theory of
Section~\ref{sec:theory} applies to any measured $f$, monotone or not. 

Low-degree agents (dark)
need a higher fraction to reach $f=\tfrac12$, high-degree agents (bright) a lower one, so the threshold
fraction $\phi^{*}(k)$ drifts downward with degree. (We define $\phi^{*}(k)$ as the fraction $x$ at which $f$
reaches $\tfrac12$: the midpoint of the response, where {sharing} becomes more likely than not.) At large degree this drift is common to all four {posts}
($\phi^{*}=0.28$--$0.34$ at $k=20$, $0.18$--$0.29$ at $k=50$ and $0.08$--$0.12$ at $k=152$); at the smallest
degrees it is not, as we discuss in Section~\ref{sec:content}.
The response is therefore neither a pure fraction rule
($\phi^{*}$ constant) nor a pure count rule. Instead, we find a fraction-like behavior with a systematic degree drift, where higher-degree agents tend to ignite at a lower fraction. {This agrees with De Marzo et al., who found that how strongly LLM agents follow a majority depends on the size of the group \cite{de2026ai}.} 

\ipar{The response depends on the content}\label{sec:content}\label{sec:watts}

The drift with degree is not the only systematic feature of the response. With the same model, the same prompt template and the same degree, the response also changes with the content of the post, and it changes most where the social signal is weakest. Consider an agent that follows three accounts, one of which has shared the post. It shares \post{gun-right} (``Everybody should have the right to own a gun.'') with probability $f(1,3)=0.82$, but each of the other three posts with probability at most $0.23$. At large degree, by contrast, the four posts are alike (Table~\ref{tab:lowk}, Section~\ref{sec:lowk-sm}).
This difference at low degree decides the collective outcome on a sparse network, where most agents follow few accounts. There, a cascade can grow from a small seed only if a single sharing contact is often enough to make an agent share, which is Watts' cascade condition \cite{watts2002simple}. On ER $z=2$ this condition nearly holds for \post{gun-right} and is far from holding for the other three posts (Section~\ref{sec:lowk-sm}). Accordingly, \post{gun-right} is the only post whose cascade on this network jumps, from a seed of $\rho_0\approx0.016$ to a final size of $\rho_\infty\approx0.73$, while the cascades of the other three grow smoothly with the seed (Fig.~\ref{fig:cascade}B).

\subsection{\texorpdfstring{Predicting the cascades requires the response measured with the degree}{Predicting the cascades requires the response measured with the degree}}
\label{sec:cascade}

\ipar{Previous studies give every agent the same response}
Measuring the degree-resolved response requires querying the model for every number of sharing contacts at every degree that occurs in the network.
Previous studies that predict the collective behavior of LLM agents from the response of single agents~\cite{de2026ai,flint2026group} measure a single response function in populations where {all agents have the same number of contacts} (a fully
connected group, in which every agent sees all the others, or pairs of agents drawn at random) and apply it to
every agent. A response that depends on the fraction alone gives an agent with 2 sharing contacts out of 10 the same probability as one with 20 out of 100; the measured response gives at most $0.01$ to the first and between $0.20$ and $0.94$ to the second, depending on the post (Fig.~\ref{fig:cascade}A). On networks, where agents differ in how many contacts they have, we show that the prediction of the cascade fails without quantifying the degree dependence. 
{This setting tests which aspects of a single agent's response must be measured to predict the collective behavior.}

\ipar{Predicting a cascade from a single-agent response}\label{sec:theory}

A measured response predicts a cascade in two ways. In Monte Carlo simulations, each LLM query is replaced by a random draw from the response, on the same networks and with the same update rule as in the LLM populations (Materials and Methods). With the empirically measured $f(m,k)$ at hand, we can go one step further: instead of performing Monte Carlo simulations of the cascades, we can predict them analytically using the tree-based mean-field theory of Gleeson and Cahalane \cite{gleeson2007seed}{, which assumes that the network is locally tree-like and that the degrees of neighboring agents are uncorrelated, conditions that our networks nearly satisfy (Section~\ref{sec:frozen-sm})}. This gives both an efficient way to compute the final cascade size on any network and analytical insight into why cascades in LLM populations ignite where they do.

The theory takes as input a response function $F(m,k)$: the probability that an agent of degree $k$ with $m$ active neighbors is itself active. The crucial step is therefore to relate the measured single-decision probability $f(m,k)$ to this response. The two are not the same, because an agent is asked again each time its exposure ($m$) grows. An agent that ends the cascade with $m$ sharing neighbors has been asked once at each exposure level $m'=1,\dots,m$, or until it shared. Since its decisions are independent, the probability that it has shared is one minus the probability that it declined every time:
\begin{equation}
  \Feff(m,k) \;=\; 1-\prod_{m'=1}^{m}\big(1-f(m',k)\big), \qquad \Feff(0,k)=0 .
  \label{eq:feff}
\end{equation}
Two properties make $\Feff$ the appropriate input to the theory. First, it depends only on the final count $m$ and the degree $k$, not on the order in which the neighbors shared. Second, it never decreases with $m$, since every further exposure is one more chance to share, regardless of the shape of $f$ (Section~\ref{sec:frozen-sm}). Equation~\eqref{eq:feff} thus links the microscopic decision we measure from the LLM to the macroscopic cascade on the network. Equivalently, each agent behaves as if it drew a hidden threshold once, from the distribution $\Feff(\cdot,k)$, and shared as soon as $m$ reached it (Section~\ref{sec:frozen-sm}). Since $\Feff$ depends on $m$ and $k$ separately, and not only on the fraction $m/k$, our model is an instance of the generalized, degree-dependent threshold model \cite{dodds2004universal,gleeson2008cascades,burkholz2018correlations}, rather than the fractional-threshold model \cite{watts2002simple}. 

The mean-field theory holds for any response $F(m,k)$ that is non-decreasing in $m$; we write it for general $F$, because the same equations will be used to quantify the error of the degree-averaged response (Section~\ref{sec:errortheory}). An agent of degree $k$ whose neighbors are each active with probability $q$ becomes active with probability $A[F](k,q)$, the average of $F(m,k)$ over the binomial number $m$ of active neighbors among the $k-1$ neighbors other than the one through which the agent is reached (Section~\ref{sec:methods-theory}). 
The probability $q$ that a randomly chosen edge leads to an active agent obeys {the self-consistency equation of Gleeson and Cahalane~\cite{gleeson2007seed}}, and its solution $q^{*}$ gives the final cascade size:
\begin{equation}
  \begin{gathered}
  q \;=\; G[F](q)\equiv\rho_0 + (1-\rho_0)\sum_k q_k\, A[F](k,q),\\[2pt]
  \rho_\infty \;=\; R[F](q^{*})\equiv\rho_0 + (1-\rho_0)\sum_k p_k\, H[F](k,q^{*}),
  \end{gathered}
  \label{eq:fixedpoint}
\end{equation}
with $p_k$ the degree distribution, $q_k=kp_k/\langle k\rangle$ the excess-degree distribution, and $H[F](k,q)$ the probability that an agent of degree $k$ is active, given that each of its $k$ neighbors is active with probability $q$ (the same binomial average, over all $k$ neighbors). We take $q^{*}$ to be the smallest fixed point of $q=G[F](q)$, obtained by iterating from $q=0$: the cascade grows from the seeds and stops at the first fixed point it reaches, even if larger ones exist. Note that $A$, $H$, $G$ and $R$ are all linear in the response $F$, a property we exploit below.

\ipar{Averaging over degrees mispredicts the cascade}\label{sec:fbar}\label{sec:mispredict}
We compare the cascade predicted with the measured response $f$ with those predicted by two degree-blind responses: (i) the average of our measured responses over the degrees of the network and (ii) the response measured in a single large group.

{The first approximation is built} from
the same measured responses{,} discarding only the information about degree, so that the comparison isolates the
effect of ignoring the degree{. It is the degree-averaged response $\fbar(x)$, the plain average, over all the degrees that occur in the network, of the measured responses read as functions of the fraction $x=m/k$ of sharing contacts (Materials and Methods), and it} gives an agent
with $m$ sharing contacts out of $k$ the probability $\fbar(m/k)$, whatever its degree. Through eq.~\eqref{eq:feff}, the cascade with $\fbar$ still depends on degree, because an agent with more contacts has more chances to share before a given fraction of them has shared; what $\fbar$ removes is the dependence of each decision on degree. In
Fig.~\ref{fig:cascade}A, $\fbar$ is the thick orange curve (shown for ER $z=5$); it disregards the spread of the
colored curves around it. We use the plain average because it is the simplest choice. 
{Weighting each degree by how many agents have it or by how often a cascade reaches it, as a measurement on randomly chosen agents or on the agents that a cascade reaches would do, does not remove the error: it is still at least $0.029$} in every combination of post and network (Section~\ref{sec:weights-sm}).

{{In Fig.~\ref{fig:cascade}, orange marks the degree average: $\fbar$ itself in panel A (for ER $z=5$), and in panels B to D the cascades with $\Feff[\fbar]$, from the mean-field theory (dashed lines) and the Monte Carlo simulations (triangles).} The theory matches the Monte Carlo {simulations} with $\fbar$, as it matches {those} with $f$, for every post and network, so the gap between the two cascades is the error of the degree-averaged response alone. We write this error as $\Delta\rho_\infty(\rho_0)=\rho_\infty[f]-\rho_\infty[\fbar]$, the final size of the cascade with $f$ minus that of the cascade with $\fbar$ at the same seed fraction $\rho_0$: it is negative where averaging overpredicts the cascade and positive where it underpredicts it.}
For three of the four {posts} --- the two migration {posts} and
\post{police-defunded}, in the first three rows of Fig.~\ref{fig:cascade} --- the degree-blind $\fbar$ ignites {every} network at a lower seed
(earlier tipping on ER $z=5$ and BA, uniformly higher $\rho_\infty$ on ER $z=2$).
The fourth {post}, \post{gun-right} (bottom row), behaves in the same way on ER $z=5$ and BA but not on ER $z=2$ {(see below)}.

The overprediction, for these three posts on every network and for \post{gun-right} on ER $z=5$ and BA, follows from how the response depends on the degree. The threshold fraction $\phi^{*}(k)$, the fraction of sharing contacts at which the response of an agent of degree $k$ reaches $\tfrac12$ (Section~\ref{sec:degree}), falls with degree (Fig.~\ref{fig:cascade}A). The degree average gives the same weight to every degree that occurs in the network, so its threshold is lower than that of the typical agent, which has few contacts, and the cascade with $\fbar$ spreads more than the cascade with $f$. {On BA, where most agents have 3 to 5 contacts but 147 different degrees occur, the degrees $k\ge10$, which only $11\%$ of the agents have, carry $94\%$ of the weight in $\fbar$, and the cascade with $\fbar$ ignites the network from a vanishingly small seed (Section~\ref{sec:cascade-sm}).} 
{The networks sit in different regimes. On ER $z=2$ the transition is continuous, except for \post{gun-right} (Section~\ref{sec:content}): the final size grows smoothly with the seed fraction. On ER $z=5$ and BA it is discontinuous: the final size stays small below the tipping seed and jumps to a global cascade there (Fig.~\ref{fig:cascade}B to D).}

The exception is \post{gun-right} on ER $z=2$. On this sparse network the cascade grows through agents with few contacts, whose response to a single sharer is high for this post (Section~\ref{sec:content}). The degree average underestimates this response: at the fraction $x=1/3$ it gives $\fbar(1/3)=0.45$, while the measured $f(1,3)$ is $0.82$. Averaging therefore underpredicts the final size, by up to $\Delta\rho_\infty\approx+0.5$, and misses the global cascade of the LLM population. On ER $z=5$ and BA the cascade grows through agents of higher degree, where the four posts respond alike, and the degree average overpredicts the cascade of \post{gun-right} as it does for the other posts. Section~\ref{sec:closedform} shows that, to first order, the error has the sign of the response deviation where the cascade samples the response.

\ipar{A large-group response mispredicts the cascade as well}
The second approximation mimics a measurement in a large group in which every agent sees all the others, as in De Marzo et al.~\cite{de2026ai}. In our measurement an agent with $k$ contacts sees $k$ others, so our largest degree, $k=155$, is the largest such group we have; we give its response to every agent. This large-group response fails, in opposite directions depending on how it is read: read at the fraction $m/k$, it ignites every network from the smallest seed; read at the count $m$, it hardly spreads beyond the seed. For every post and network, with either reading, its error, the mean of $|\Delta\rho_\infty|$ over the seed fractions, is at least as large as that of $\fbar$, with one exception, \post{gun-right} on ER $z=2$ read at the fraction, where the agreement is accidental. Because it gives every agent the response of a degree that almost none of them has, it fails even when every agent has the same degree, where $\fbar$ is exact (mean-field theory, Section~\ref{sec:transfer-sm}).

\ipar{The response measured with the degree reproduces the LLM populations}\label{sec:mc}

The Monte Carlo simulations with the measured $f(m,k)$ (black points in Fig.~\ref{fig:cascade}B--D) closely reproduce the cascades of the full LLM populations (blue squares). This agreement is expected: each agent's decision is binary, independent and memoryless, and depends only on the post, $m$ and $k$. It is therefore a Bernoulli trial fully characterized by $f(m,k)$, which captures not just the mean of the sharing behavior but its entire distribution. Cascades on networks of LLM agents can therefore be predicted from single-agent measurements alone, but only if the measurement records every variable the agent responds to: here the degree as well as the number of sharing contacts, which the degree-blind responses above leave out.

{The LLM population follows the Monte Carlo simulations with $f$ on every network, except that \post{police-defunded} and, to a lesser extent, \post{migration-must-be-limited} spread slightly earlier than their responses predict.} On the two ER networks, the LLM population of \post{police-defunded} lies closer to the cascade with $\fbar$ than to that with $f$.

Fed with $F=\Feff$, the theory reproduces the behavior of the full LLM populations on all networks and for all posts with good accuracy (black lines in Fig.~\ref{fig:cascade}B--D). The deviations close to the tipping points, especially on scale-free networks, are finite-size effects that smooth out the transition.

\subsection{\texorpdfstring{The error of the degree average grows with network heterogeneity}{The error of the degree average grows with network heterogeneity}}
\label{sec:r4}

The error of the degree average differs widely between posts and networks (Fig.~\ref{fig:cascade}). This section shows what sets its size: how much the response depends on degree, and how heterogeneous the network is.

\begin{figure}[tbp]
  \centering
  \includegraphics[width=0.49\textwidth]{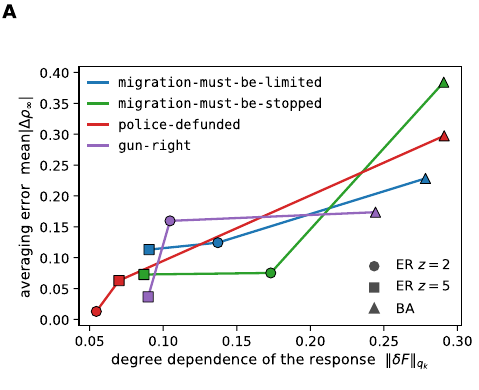}\hfill
  \includegraphics[width=0.49\textwidth]{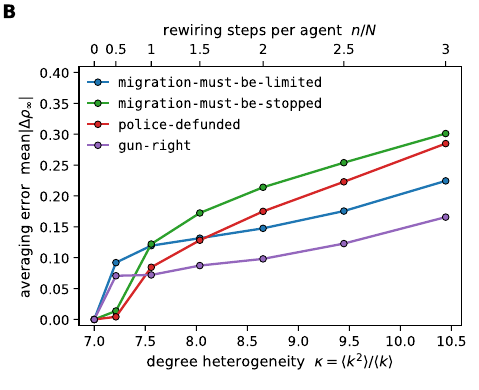}
  \caption{{\textbf{The error of the degree average grows with the degree dependence of the response and with the heterogeneity of the network.} (\textbf{A}) Averaging error mean$|\Delta\rho_\infty|$ against the degree dependence of the response $\|\delta F\|_{q_k}$, from the Monte Carlo {simulations} (25 seed fractions). Color: post; marker: network; the lines join the three networks of each post. The error grows along the line of every post and is largest on BA. (\textbf{B}) Averaging error against the degree heterogeneity $\kappa=\langle k^2\rangle/\langle k\rangle$ at fixed mean degree $\langle k\rangle=7$, from the mean-field theory (601 seed fractions). The networks are obtained by rewiring a regular graph, on which $\kappa=7$ (top axis: rewiring steps per agent, $n/N$). The error is zero on the regular graph and grows with heterogeneity for every post.}}
  \label{fig:error}
\end{figure}

\ipar{The error grows with the degree-dependence of the response}
The error of the degree average should grow with how far the degree-resolved response lies from it. At each count $m$ and degree $k$, this distance is $\delta f(m,k)=f(m,k)-\fbar(m/k)$, the vertical gap between a colored curve and the orange curve in Fig.~\ref{fig:cascade}A. The cascade, however, depends on the response through the effective response $\Feff$ of eq.~\eqref{eq:feff}, so we take the gap there, $\delta F=\Feff[f]-\Feff[\fbar]$; to first order, the cascade error is a weighted sum of $\delta F$ (Section~\ref{sec:linear}). To compare posts and networks, we condense $\delta F$ into one number, the degree dependence of the response $\|\delta F\|_{q_k}$. It is the root-mean-square of $\delta F$ over $m$ and $k$, with each degree weighted by $q_k$, the probability that a neighbor of an agent has degree $k$, because a cascade reaches agents through their neighbors (Section~\ref{sec:gold}; Section~\ref{sec:choices-sm} explains this choice). Figure~\ref{fig:error}A plots the averaging error mean$|\Delta\rho_\infty|$ against this number for every post and network, with one line per post joining the three networks.

The error grows with the degree dependence of the response along the line of every post, and every post has its largest error on BA, where that dependence is largest (mean$|\Delta\rho_\infty|=0.17$--$0.38$). The points do not, however, fall on one curve, as they would if $\|\delta F\|_{q_k}$ alone determined the error. On ER $z=5$, for instance, three posts have almost the same degree dependence ($\|\delta F\|_{q_k}$ between $0.087$ and $0.090$), but their errors differ by a factor of three: $0.037$ for \post{gun-right}, $0.073$ for \post{migration-must-be-stopped} and $0.113$ for \post{migration-must-be-limited}. The error depends not only on how much the response depends on degree, but also on where, at which counts and degrees, because the cascade samples the response only at some of them (Section~\ref{sec:linear}). It therefore cannot be forecast from $\|\delta F\|_{q_k}$, or from any other single number that summarizes the response.

\ipar{Heterogeneity at fixed mean degree}\label{sec:hetero}
{We now test how the error depends on network heterogeneity. The networks of Fig.~\ref{fig:error}A differ in both mean degree and heterogeneity. To separate the two, we start from a regular graph with $\langle k\rangle=7$ and rewire it step by step~\cite{ojer2025social}, which makes the degrees more unequal at fixed mean degree ($n/N$ denotes the number of rewiring steps per agent), and we compute the error with the mean-field theory (Materials and Methods), 
{which reproduces the Monte Carlo simulations for both responses on every network (Fig.~\ref{fig:cascade}).} Heterogeneity is measured by $\kappa=\langle k^2\rangle/\langle k\rangle$.}
On the regular graph ($\kappa=7$) the error is zero {(Fig.~\ref{fig:error}B)}, because a
network with a single degree leaves nothing to average over. As the rewiring makes the network more heterogeneous,
the error grows for all four {posts}, reaching $0.17$--$0.30$ at the largest heterogeneity ($\kappa\approx10.4$,
$n/N=3$). The growth is monotonic for every {post}, although slow for \post{gun-right} between $n/N=0.5$ and $1$. Section~\ref{sec:closedform} explains this growth analytically.

At $\langle k\rangle=6$, 9 and 12 the error also grows with heterogeneity (Fig.~\ref{fig:sm_hetero}): the rank correlation between the error and $\kappa$ is positive in all sixteen sweeps (four posts at four mean degrees), between $0.89$ and $1.00$. We {draw no conclusion} about how the error depends on the mean degree
itself, because that dependence comes from how $\fbar$ is built: the plain average gives every degree that occurs the same weight, and a denser network spans more degrees, so its average contains more of the permissive high-degree responses. 
{With the averages that weight each degree by how many agents have it or by how often a cascade reaches it, the trend reverses} (Section~\ref{sec:meandegree-sm}).

{Heterogeneity therefore} controls the cost of degree-averaging: the error vanishes on a homogeneous
network and grows as the degrees become more unequal, for every {post} and every mean degree we tested.

\subsection{\texorpdfstring{{Predicting the averaging error requires measuring the degree dependence}}{Predicting the averaging error requires measuring the degree dependence}}
\label{sec:closedform}

\ipar{The degree dependence cannot be corrected for}

{If the error of the degree-averaged response could be predicted from the network and from a single number summarizing how much the response depends on degree, a degree-blind response corrected by a known factor would suffice, and the degree dependence could be left unmeasured. We test this in two steps. First, we derive the error analytically: we expand the cascade theory of Section~\ref{sec:theory}~\cite{gleeson2007seed} around the degree-averaged response, which gives, to first order and in closed form, the error of any degree-blind response. Second, we test whether this error factorizes into network statistics times a single number for the response. 
It does not: the error depends on the full shape of the degree dependence, which must therefore be measured, not corrected for.
}

\ipar{The error is a weighted sum of the response deviation}\label{sec:errortheory}\label{sec:linear}
Write the two responses
that feed the theory as $F=\Feff[f]$ ({degree-resolved}) and $\bar F=\Feff[\fbar]$ ({degree-averaged}), and their
difference $\delta F\equiv F-\bar F$. {We expand the error around the solution with $\fbar$, the smallest fixed point $q_0$ of $q_0=G[\bar F](q_0)$. Because the equations of the theory are linear in the response, to first order in $\delta F$ the error in the final cascade size is a weighted sum of the response deviation (Section~\ref{sec:methods-error}):}
\begin{equation}
  \begin{gathered}
  \Delta\rho_\infty \;=\; \sum_{m,k} W(m,k)\,\delta F(m,k),\\[2pt]
  W(m,k)=(1-\rho_0)\Big[\,p_k\,b_m(k,q_0) \;+\; \chi\,R'[\bar F](q_0)\,q_k\,b_m(k{-}1,q_0)\Big],
  \end{gathered}
  \label{eq:linear}
\end{equation}
where $b_m(n,q)=\binom{n}{m}q^m(1-q)^{n-m}$ is the binomial {probability of $m$ active neighbors out of $n$, $R'$ and $G'$ are the derivatives of $R$ and $G$ with respect to $q$, and $\chi=1/(1-G'[\bar F](q_0))$ is the} susceptibility. The susceptibility measures how strongly the cascade amplifies a small change in the response: a link that leads to an extra active agent makes, on average, $G'$ further links do so, and $\chi=1+G'+G'^2+\dots$ sums this chain. {The first term is the direct change in the final size; the second is the change in how activity propagates along edges, amplified by $\chi$.} 
{We tested this first-order expansion against the Monte Carlo simulations (Section~\ref{sec:newtheory-sm}). It holds where the two cascades stay close: where both grow smoothly with the seed fraction, as on ER $z=2$, and where both jump at nearby tipping seeds, as on ER $z=5$, for which it gives the shift of the tipping seed. There it reproduces the measured error within $20\%$, except in the case with the largest error, \post{migration-must-be-limited} on ER $z=2$, which it overestimates by about a third: as expected for a first-order expansion, the overestimate grows with the size of the error. It fails where, over a wide range of seeds, one cascade has spread to most of the network and the other has not: on BA, where the cascade with $\fbar$ spreads from a vanishingly small seed and the cascade with $f$ only from $0.04$--$0.13$, and for \post{gun-right} on ER $z=2$, whose cascade with $f$ jumps at $\rho_0\approx0.016$.}

{Three properties of the weight explain the errors of Sections~\ref{sec:cascade} and~\ref{sec:r4}. First, $W$ is never negative, so the error has the sign of $\delta F$ where $W$ is large, that is, where the cascade samples the response: when the degree-averaged response is too permissive there ($\delta F<0$), the average overpredicts, as for three of our four posts on every network and for \post{gun-right} on all but the sparsest one. Because this sign depends on the post, no single correction factor can account for it. Second, $W$ is a binomial peaked at $m\approx kq_0$, the counts the cascade samples, and its propagated part is tilted toward hubs by $q_k$, so the error grows with degree heterogeneity. Third, $\chi$ diverges at the tipping point, where $G'[\bar F](q_0)\to1$, so the deviation is amplified most there. The same expansion gives the shift $\Delta\rho_0^{*}=\rho_0^{*}[f]-\rho_0^{*}[\fbar]$ of the tipping seed as another weighted sum of $\delta F$ (Section~\ref{sec:methods-error}); it is positive in all our discontinuous cases, and the degree-blind response underestimates how hard the network is to ignite.}

\ipar{The error does not factorize into network and response}\label{sec:factorization}
The error could be predicted without measuring the degree dependence if it factorized into a property of the network times a single number for the response, $\mathrm{error}\approx A_{\rm net}(\langle k\rangle,c_v^2)\times g$. Here $A_{\rm net}$ is an amplification set by the network alone, through its mean degree $\langle k\rangle$ and the squared coefficient of variation of its degrees, $c_v^2=\langle k^2\rangle/\langle k\rangle^2-1$, and $g$ is a single number for how strongly the response depends on degree. A typical value of $g$ could then replace the measurement, and degree averaging would be a controlled approximation.

The test needs a number $g$ that describes the response alone, because it asks whether the error separates into a part set by the network and a part set by the response. The degree dependence $\|\delta F\|_{q_k}$ of Fig.~\ref{fig:error}A does not qualify: it is computed from the degree average of each network and weighted by the $q_k$ of that network, so it changes from network to network. We therefore take $g$ to be the root-mean-square departure of the measured response from a pure fraction rule, which depends only on the post (Section~\ref{sec:methods-error}).

We test the factorization on the heterogeneity sweeps of Section~\ref{sec:hetero} (Fig.~\ref{fig:sm_hetero}). If it held, the ratio error$/g$ would be the same for all posts on a given network. Instead, it differs between posts by a factor of about two (Section~\ref{sec:methods-error}), although $g$ hardly differs between them ($0.13$--$0.15$). The reason is the weight $W$ of eq.~\eqref{eq:linear}: it gives the deviation of the response the most weight where the cascade samples it, at counts near $kq_0$, at degrees weighted toward hubs, and near the tipping point. Two responses with the same $g$ can deviate in different places, and a single number cannot tell them apart. The network thus sets the scale of the error, but not its size for a given post. Degree averaging is therefore not a controlled approximation on heterogeneous networks, and the degree-resolved response $f(m,k)$ must be measured rather than summarized.

\section{\texorpdfstring{Discussion}{Discussion}}
\label{sec:discussion}
\label{sec:next}
The collective behavior of networked LLM agents can be predicted from measurements of a single agent, provided the agent is measured in the local context in which it acts. In our networks, each agent decides whether to share a post after seeing how many of the accounts it follows have shared it. {Populations of such agents produce information cascades whose reach depends on the network and on the post. How an agent responds depends on how many contacts it has: the more contacts it has, the lower the fraction of them that must have shared before sharing becomes more likely than not. The post matters mainly when the agent has few contacts.} The response of a single agent to $m$ sharing contacts out of $k$, measured at every degree and fed into the mean-field theory of cascades, predicts the cascades of the full LLM simulations, including whether they spread and the seed fraction at which they tip. Measured without this context, averaged over degree or taken from a single large group, the response mispredicts the cascades, increasingly so on heterogeneous networks and in a direction set by the content of the post. 

These results bear on what a measurement of a single agent must capture. Predicting the collective behavior of LLM populations from the behavior of single agents has succeeded in several settings \cite{de2026ai,flint2026group,el2026physics}, most of them well-mixed groups in which all agents face the same situation. Where agents face different situations, the prediction holds only if the measurement resolves them, because the response is a property of the agent in its context, not of the agent alone. On a social network the context includes the number of contacts. Our agents follow neither of the two rules assumed by classic threshold models, a fixed fraction or a fixed number of sharing contacts \cite{granovetter1978threshold,watts2002simple,dodds2004universal,centola2007complex}: their threshold fraction falls with their degree, in line with the weakening of majority-following with group size in well-mixed groups \cite{de2026ai}. {In human populations the evidence is mixed: the adoption thresholds of Skype users, measured as fractions of their contacts, appear not to depend on degree~\cite{karsai2016local}, whereas on Twitter highly connected users are less likely to share any given message, because their attention is divided among more incoming messages~\cite{hodas2012visibility,gomez2014quantifying}.} A response measured out of context mispredicts the collective outcome, in line with the observation that the collective biases of LLM populations are not easily deduced from agents in isolation \cite{ashery2025emergent}. The network thus shapes the collective outcome not only through who interacts with whom \cite{passaro2026network}, but also through the local context it creates for each agent, which changes how the agent responds.

Measuring the local response properly, and therefore predicting the collective behavior, is not trivial. The response must be known at the degrees that occur in the network, but the structure of real social networks is rarely known in full, and it changes over time, as accounts join and leave and links form and dissolve~\cite{kossinets2006empirical,gross2008adaptive}. A measurement must therefore anticipate degrees not yet observed. Covering every degree is costly: an agent with $k$ contacts can see $k$ nonzero counts of sharing contacts, each needing its own queries, so the cost grows with the square of the largest degree; up to $k=155$ we made about two million queries per post. Unmeasured degrees must be interpolated or extrapolated, and whether the response of an LLM agent is regular enough for this is open. On heterogeneous networks the degrees hardest to cover, those of agents with many contacts, also carry more weight than their share of the population: in the error derived in Section~\ref{sec:closedform}, the part that propagates along edges weights the response of each degree by $q_k$, the probability that a neighbor has that degree. A measurement made for one network thus carries over to another only if it covers that network's degrees.

{We derived the error of the degree-averaged response analytically. To first order, it is a sum over the pairs $(m,k)$ of the deviation of the response from its degree average, each weighted by a factor set by the network and by the cascade with the degree-averaged response (eq.~\eqref{eq:linear}).} This form of the error has a practical consequence: a prediction made with a degree-blind response cannot be corrected by a factor computed from the statistics of the network. The error does not factorize into a property of the network times a single number for the response, because the weight selects where the deviation matters, at the pairs $(m,k)$ that the cascade samples, and posts whose responses are equally degree-dependent differ there. On the same networks, the error per unit of degree dependence varies between posts by a factor of about two (Section~\ref{sec:closedform}). The degree dependence of the response therefore has to be measured, not corrected for. The weight also shows where the measurement must be most precise. It is amplified near the tipping point, where the susceptibility $\chi$ diverges, so whether a post spreads is the prediction most sensitive to errors in the measured response.

Our conclusions rest on one language model, one prompt and four posts, and our agents have only a minimal persona and no memory. 
{The response to \post{gun-right} differs from that to the other three posts at low degree}{, which sets its cascade on the sparse network apart (Section~\ref{sec:content}), but} whether this difference survives rewordings of the post has not been tested. The LLM population follows its measured response closely, but for \post{police-defunded} and, to a lesser extent, for \post{migration-must-be-limited} it spreads slightly earlier than the response predicts (Section~\ref{sec:cascade}). The expansion of the error is first order: on ER $z=2$ it overestimates the largest error by up to about a third, for \post{migration-must-be-limited} (Section~\ref{sec:newtheory-sm}), and the expansion fails where {the two cascades do not stay close}, on BA and for \post{gun-right} on ER $z=2$, where the error must be computed from the two mean-field solutions. Finally, our networks are synthetic and locally tree-like, and on clustered networks the tree theory loses accuracy. 

\label{sec:future}
Our agents see only how many of their contacts have shared, so the measured response describes them completely, and the full LLM simulations follow it closely. Agents deployed in practice carry more: a more complex persona or an ideology \cite{argyle2023out,santurkar2023whose,piao2025emergence}, a memory of past interactions \cite{park2023generative,ashery2025emergent,flint2026group}, and the content of what their contacts shared and the identity of who shared it \cite{chuang2024simulating,nohr2026peer}. Each of these makes the response depend on more than $(m,k)$. A persona makes the response differ from agent to agent; memory makes the successive decisions of an agent depend on each other, so that its effective response no longer follows from $f$ by eq.~\eqref{eq:feff}; and the arguments attached to a message can sway an agent more than the identity of its sender \cite{nohr2026peer}. For such agents the response can be refined step by step, from a single $f$ to one response for each persona and then to a policy that depends on the history of the agent, until the predicted and the simulated collective behavior agree. The gap that remains at each step measures how predictable a given design of agents is, and the step at which it closes identifies the minimal description of the agent that the collective behavior requires.

A measured response also indicates where to intervene. The theory extends directly to seed fractions that depend on degree, so it predicts how much the tipping seed falls when the first sharers are hubs, a placement with which far fewer committed agents overturn a convention in networked LLM populations \cite{passaro2026network}. The weight $W$ shows whose response matters most for the collective outcome, on heterogeneous networks that of the hubs, and therefore which agents an audit or a change of instructions should target first. The way a platform displays social proof, as a count, as a fraction or not at all, changes the response itself, and the same pipeline can measure how each choice changes the collective outcome. 

The approach extends to clustered networks, for which cascade theories exist \cite{hackett2011cascades}, to networks that co-evolve with the behavior of the agents \cite{gross2008adaptive,piao2025emergence}, and to reversible dynamics such as consensus and polarization \cite{castellano2009statistical,baumann2020modeling}, whose mean-field theories also take a response resolved by degree \cite{gleeson2013binary}. Because a measured response predicts the collective outcome on any network whose degrees it covers, without further queries to the model, responses measured for each model, once their predictions have been tested on real-world social networks, could anticipate the collective behavior of a population of agents before it is deployed \cite{hammond2025multi}, including populations that mix models, which need one response per model \cite{nohr2026peer}.

\section{\texorpdfstring{{Materials and Methods}}{Materials and Methods}}
\label{sec:methods}

\subsection{\texorpdfstring{{The LLM agents and the measurement of the response}}{The LLM agents and the measurement of the response}}
\label{sec:methods-prompt}\label{sec:methods-kernel}
{Every query, in the measurement of the response and in the full LLM simulations, uses the following prompt, in which \texttt{\{m\}} and \texttt{\{k\}} {are replaced by} the number of the agent's neighbors who have shared the post and {by its degree}:}

\definecolor{shadecolor}{RGB}{242,242,242}
\begin{shaded}\small
\begin{verbatim}
You are a user on a social media platform that tries to maximize engagement.
A post by another user is shown below.

Post: "..."

The post is supported by {m} out of {k} of the peers you follow.

Do you support the post to reshare it? Answer with exactly one word: YES or NO.
\end{verbatim}
\end{shaded}

{Each element of the prompt serves the measurement. The persona, a user of a social media platform that tries to maximize engagement, gives the agent a reason to share; the count ``$m$ out of $k$'' lets it respond to the number of sharers, to their fraction or to both, so that the response can depend on degree; and the one-word answer makes each query a binary decision. Because queries are stateless and show only this count, the response depends on $(m,k)$ alone, as the mean-field theory assumes.} {The prompt treats sharing as support: it describes the contacts who have shared the post as supporting it.} 

{{We measure the response $f(m,k)$ of Section~\ref{sec:model-def} for the LLM \texttt{mistral:7b} (via Ollama, $T=0.3$) and each post.} For each pair $(m,k)$ we make an independent set of 200 calls, with the same prompt
as in the full LLM simulations, and $f(m,k)$ is the fraction of YES answers; its standard error,
$\sqrt{f(1-f)/200}$, is at most $0.035$. If the model answers neither YES nor NO but declines to take
a position, the query is repeated, and after five such answers in a row the decision is drawn at random, YES or NO with equal probability. The same rule applies in the full LLM simulations.} {We draw the decision at random, rather than counting a non-answer as NO, so as not to bias the response toward sharing or toward not sharing.}

\subsection{\texorpdfstring{Full LLM and Monte Carlo simulations}{Full LLM and Monte Carlo simulations}}
\label{sec:methods-sims}
We run the exposure-triggered cascade on three kinds of network, all with $N=1000$ agents: Erd\H{o}s--R\'enyi
(ER) graphs \cite{erdos1959random} with mean degree $z=2$ and $z=5$, and Barab\'asi--Albert (BA) graphs \cite{barabasi1999emergence}{, in which every new agent links to three existing ones}.
On ER, $z=2$ is sparse enough that the final size grows continuously with the seed,
while $z=5$ is dense enough for a discontinuous jump, so the two cover both kinds of transition. BA
has a mean degree close to 6, near that of ER $z=5$, so going from ER $z=5$ to BA changes mainly the
heterogeneity of the degrees. For each network we generate 500 graphs. Each Monte Carlo cascade starts from a
fraction $\rho_0$ of randomly chosen seed agents, for 25 values of $\rho_0$ between $0.005$ and $0.30$; this
window extends beyond the largest tipping seed of the {degree-resolved response} ($0.21$), so that every jump is captured. For
each seed fraction we run 500 cascades, on the 500 graphs, and average their final size. The {LLM population}, with an LLM query at every exposure, runs on the first 100 of these graphs: 100
cascades at each of 12 seed fractions between $0.02$ and $0.26$ (9 seed fractions, up to $0.19$, for
\post{gun-right} on BA). {The median standard error of the average final size is $0.0003$ for the Monte Carlo simulations and $0.001$ for the LLM population; the largest, $0.02$ and $0.04$, occur at the jumps, where single runs either ignite or not.} Two rules complete the exposure-triggered cascade of Section~\ref{sec:setup}. If an agent's exposure grows by more than one neighbor within a single round, it considers each new exposure level in turn and stops as soon as it decides to share. Sharing is irreversible: agents remain sharers for the rest of the cascade once they have shared the post. The effective response $\Feff$ (Section~\ref{sec:theory}) and the hidden-threshold picture of Section~\ref{sec:frozen-sm} rely on both rules.

\subsection{\texorpdfstring{{The degree-averaged response and the error observables}}{The degree-averaged response and the error observables}}
\label{sec:gold}
{\emph{The degree-averaged response.}} The measured response of each degree $k$ is read as a
function of the fraction $x=m/k$ of sharing contacts, and interpolated linearly between the points
$x=0,1/k,\dots,1$. The degree average $\fbar(x)$ is the plain average of these curves over all the degrees that occur
in the network, each degree counting once, however many agents have it. On a
homogeneous network it reduces to the single measured curve.

\emph{The averaging error} (vertical axis {of Fig.~\ref{fig:error}A}) is mean$|\Delta\rho_\infty|$: the absolute difference between the final
cascade size obtained with the degree-resolved {response} and with the degree average $\fbar$, averaged over the 25
seed fractions $\rho_0$ between $0.005$ and $0.30$ of the Monte Carlo simulations. We take the absolute value so
that errors of opposite sign at different seeds do not cancel, and we use the same window of seeds for every {post}
and network so that all cases are compared on the same footing.

\emph{The degree-dependence of the response} (horizontal axis of Fig.~\ref{fig:error}A) is
\begin{equation}
  \|\delta F\|_{q_k}=\Big[\sum_k q_k\,\frac{1}{k+1}\sum_{m=0}^{k}\delta F(m,k)^2\Big]^{1/2},
  \qquad \delta F(m,k)=\Feff[f](m,k)-\Feff[\fbar](m,k),
  \label{eq:dFnorm}
\end{equation}
where $q_k=k\,p_k/\langle k\rangle$ is the
probability that a randomly chosen neighbor of an agent has degree $k$ ($p_k$ being the fraction of agents with
degree $k$). {We use the effective response $\Feff$ because it is the one that enters the theory, and the weights $q_k$ because the propagated part of the error, which dominates near the tipping point, sums over degrees with these weights; Section~\ref{sec:choices-sm} shows what the weights $p_k$ would change.}

\subsection{\texorpdfstring{{Networks of controlled heterogeneity}}{Networks of controlled heterogeneity}}
\label{sec:methods-rewiring}
{To vary heterogeneity at fixed mean degree, we start} from a regular graph in which
every agent has $\langle k\rangle$ contacts and rewire it step by step~\cite{ojer2025social}. At each step an agent $i$ with more than
$k_{\min}=3$ contacts is chosen at random, and the link to its least-connected neighbor is moved to an agent $j$
chosen anywhere in the network, with probability proportional to its degree, among the agents with at least as many
contacts as $i$. Each step keeps the number of agents and of links, and hence the mean degree, fixed, while making
the degrees more unequal: well-connected agents gain links at the expense of less-connected ones. We take
snapshots after $n/N=0,0.5,\dots,3$ steps per agent, with 100 graphs of $N=1000$ agents each{, for $\langle k\rangle=7$ (Fig.~\ref{fig:error}B) and for $\langle k\rangle=6$, 9 and 12 (Fig.~\ref{fig:sm_hetero})}.

{We measure heterogeneity by $\kappa=\langle k^2\rangle/\langle k\rangle$: $\kappa-1=\sum_k q_k(k-1)$ is the mean number of further contacts of an agent reached by following a link, the quantity that controls propagation in the theory, and at fixed mean degree $\kappa$ carries the same information as $c_v^2=\langle k^2\rangle/\langle k\rangle^2-1$. The errors are computed with the mean-field theory, which reproduced the Monte Carlo simulations in all twelve post--network cases of Fig.~\ref{fig:cascade} and lets us scan many networks without new simulations; they are averaged over 601 seed fractions in the window $0.005\le\rho_0\le0.30$ (Section~\ref{sec:choices-sm}).}

\section*{Acknowledgments}
{\sloppy F.B. acknowledges support from a University of Pennsylvania MindCORE fellowship. M.S. acknowledges support from Grant No.~RYC2022-037932-I, funded by MCIN/\allowbreak AEI/\allowbreak 10.13039/\allowbreak 501100011033 and FSE+, and Grant No.~CNS2023-144156, funded by MICIU/\allowbreak AEI/\allowbreak 10.13039/\allowbreak 501100011033 and the European Union NextGenerationEU/\allowbreak PRTR.\par}

An AI assistant (Claude, Anthropic) was used to help edit the text and to write analysis and plotting code. The authors checked all of its output and take full responsibility for the content.

\bibliographystyle{unsrt}
\bibliography{references}

\clearpage
\section*{Supplementary Materials}
\label{sec:sm}
\setcounter{subsection}{0}\setcounter{figure}{0}\setcounter{table}{0}\setcounter{equation}{0}
\renewcommand{\thesubsection}{S\arabic{subsection}}\renewcommand{\theHsubsection}{S\arabic{subsection}}
\renewcommand{\thefigure}{S\arabic{figure}}\renewcommand{\theHfigure}{S\arabic{figure}}
\renewcommand{\thetable}{S\arabic{table}}\renewcommand{\theHtable}{S\arabic{table}}
\renewcommand{\theequation}{S\arabic{equation}}\renewcommand{\theHequation}{S\arabic{equation}}

\subsection{\texorpdfstring{The response to a single sharer decides whether a small seed spreads on a sparse network}{The response to a single sharer decides whether a small seed spreads on a sparse network}}
\label{sec:lowk-sm}
On a sparse network, a cascade can grow from a small seed only if one sharing contact is often enough to make an agent with few contacts share (Section~\ref{sec:content}). This section gives that response for the four posts and evaluates Watts' cascade condition with it.

{Table~\ref{tab:lowk} compares the four posts where they differ most: the response of an agent that follows only
two or three accounts, one of which has shared.}

\begin{table}[htbp]
\centering
\small
\begin{tabular}{p{6.2cm}ccc}
\hline
post & $f(1,2)$ & $f(1,3)$ & $\phi^{*}(20)$\\
\hline
``Migration must be limited.''                   & $0.41$ & $0.20$ & $0.29$\\
``Migration must be stopped.''                   & $0.04$ & $0.02$ & $0.34$\\
``The police should be defunded.''               & $0.69$ & $0.23$ & $0.32$\\
``Everybody should have the right to own a gun.'' & $\mathbf{1.00}$ & $\mathbf{0.82}$ & $0.28$\\
\hline
\end{tabular}
\caption{\textbf{Where the four posts differ.} $f(1,k)$: probability of sharing when one of the $k$ accounts followed
has shared (measurement uncertainty $\le0.035$). $\phi^{*}(20)$: threshold fraction at $k=20$, a large degree,
inside the measured range, at which the four posts agree.}
\label{tab:lowk}
\end{table}

{Watts' cascade condition \cite{watts2002simple} makes the consequence quantitative.} $G'(0)=\sum_k q_k\,(k-1)\,F(1,k)$ is the mean number of further activations that a newly active {agent} triggers through neighbors that a single active contact suffices to switch on; above~1 a vanishing seed
spreads globally. On ER $z=2$, where degrees 2--4 carry $72\%$ of $q_k$, it is $0.79$ for \post{gun-right}
against $0.03$--$0.33$ for the other posts. The value for \post{gun-right} is close enough to~1 that a seed of $\rho_0\approx0.016$ tips the
network.  

\subsection{\texorpdfstring{Each agent behaves as if it had a hidden threshold}{Each agent behaves as if it had a hidden threshold}}
\label{sec:frozen-sm}
The mean-field theory of Section~\ref{sec:theory} requires a response $F(m,k)$ that does not decrease with $m$ and a final state that does not depend on the order in which agents update. This section shows that our protocol, in which an agent is asked again at each new exposure, meets both requirements, because it is equivalent to a model in which each agent draws a fixed threshold once.

Let us consider an agent of degree $k$. It is asked at $m=1,2,3,\dots$ in that order, each time with an independent YES
probability $f(m,k)$, so the probability that it has said NO to all of the first $m$ queries is
\begin{equation}
  S(m,k)=\prod_{j=1}^{m}\bigl[1-f(j,k)\bigr],\qquad S(0,k)=1 ,
\end{equation}
and $\Feff(m,k)=1-S(m,k)$ (eq.~\eqref{eq:feff}). $\Feff(\cdot,k)$ is non-decreasing, with $\Feff(0,k)=0$ and
$\Feff(k,k)\le1$: it is the cumulative distribution of an integer random variable $M\in\{1,\dots,k\}\cup\{\infty\}$, which is
the level at which the agent first says YES.

The same process can be described in a second way. Suppose the agent draws a threshold $M$ once, from the distribution $\Feff(\cdot,k)$, keeps it fixed, and shares as soon as $m$ reaches $M$. At each level $m$, its probability of sharing, given that it has not shared before, is then
\begin{equation}
  \Pr(\text{activate}\mid\text{not yet})=\frac{\Feff(m,k)-\Feff(m-1,k)}{1-\Feff(m-1,k)}
  =\frac{S(m-1,k)-S(m,k)}{S(m-1,k)}=f(m,k),
\end{equation}
as in the first description. The two descriptions are therefore the same process: $f(m,k)$ is the probability of sharing at level $m$ given no earlier sharing, and $\Feff$ is the distribution of the threshold. A smooth $f$ means thresholds spread over many levels; a step-like $f$ means a threshold common to all agents.

Two consequences follow. (a) $\Feff$ is monotone in $m$ by construction, whatever $f$ looks like{, including the non-monotonic $f$ we measure for some high-degree agents. Any irreversible cascade requires this property.} The condition $\Feff(0,k)=0$ is part of the model: an agent none of whose neighbors has shared never sees the post and is never asked. (b) Once the
thresholds are frozen, the final active set is the closure of the seed set under the rule $m\ge M$ and does not depend
on the update order: synchronous rounds and an asynchronous queue give the same distribution of final states, and
only the time course differs. The per-level querying of the protocol is what guarantees this: an agent whose $m$
jumps by three gets the same three chances it would have had if its three neighbors had shared in separate
rounds.

Gleeson and Cahalane list three conditions for their tree calculation to describe the process on a random graph:
the network is locally tree-like, each agent changes state at most once, and the final state does not depend on
the update order. Under our protocol the second condition holds because sharing is irreversible, and the third is
consequence (b).
The first is an assumption about the network, and the theory makes a second one, implicit in writing it with the degree distributions $p_k$ and $q_k$ alone: the degrees of neighboring agents are uncorrelated. Both hold for configuration-model graphs, such as our Erd\H{o}s--R\'enyi networks, whose degree assortativity, the correlation between the degrees at the two ends of a link, is zero on average. Our Barab\'asi--Albert networks are slightly disassortative (degree assortativity $-0.08$ on average), and the theory still reproduces the Monte Carlo simulations on them (Fig.~\ref{fig:cascade}). Clustered networks violate the first assumption.

\subsection{\texorpdfstring{{Mean-field theory of the cascade}}{Mean-field theory of the cascade}}
\label{sec:methods-theory}
The main text gives the fixed point of the theory, eq.~\eqref{eq:fixedpoint}, but describes only in words the two binomial sums of the response that enter it. This section writes them out, so that the final size can be computed from any response $F(m,k)$.

{The theory we use to analytically describe the information cascades is the one of Gleeson and Cahalane \cite{gleeson2007seed}. An agent of degree $k$, whose other neighbors are each active with probability $q$, becomes active with probability}
\begin{equation}
  A[F](k,q) \;=\; \sum_{m=0}^{k-1} \binom{k-1}{m} q^{m}(1-q)^{k-1-m}\, F(m,k),
  \label{eq:Akq}
\end{equation}
where the sum runs over the $k-1$ neighbors other than the one through which the {agent} is reached{; an agent of degree $k$ whose $k$ neighbors are each active with probability $q$ is active with probability}
\begin{equation}
   H[F](k,q)=\sum_{m=0}^{k}\binom{k}{m}q^m(1-q)^{k-m}F(m,k) .
   \label{eq:H}
\end{equation}

With eq.~\eqref{eq:fixedpoint} of the main text, these two sums give the final size of the cascade.

\subsection{\texorpdfstring{Comparing different degree averages}{Comparing different degree averages}}
\label{sec:weights-sm}
A degree-blind measurement on randomly chosen agents would weight each degree by how many agents have it, and one on the agents that a cascade reaches by how often the cascade reaches it, while our degree average $\fbar$ is the plain average of the measured response curves over the degrees present in the network, each degree counting once (Section~\ref{sec:fbar}). 
This section shows that weighting the degrees does not remove the error.

With $f(x;k)$ the measured response of degree $k$ read as a function of the fraction $x=m/k$ (Materials and Methods), the neighbor-weighted average is $\fbar_q(x)=\sum_k q_k\,f(x;k)\big/\sum_k q_k$, where $q_k=k\,p_k/\langle k\rangle$ is the probability that a randomly chosen neighbor of an agent has degree $k$ and $p_k$ is the fraction of agents with degree $k$. 
It is the response of the agents that a cascade reaches, with the weights of the mean-field equations. Weighting each degree instead by how many agents have it, with $p_k$ in place of $q_k$, gives the agent-weighted average $\fbar_p$, the response of a randomly chosen agent. 
We compute the cascades of all three averages with the mean-field theory, which reproduces the Monte Carlo simulations for both the degree-resolved and the degree-averaged response on every network (Fig.~\ref{fig:cascade}).

\emph{No degree average removes the error.} The plain average leaves an error of at least $0.03$ in eleven of the twelve post--network combinations; the exception is \post{police-defunded} on ER $z=2$, with $0.013$ (Table~\ref{tab:weights}). An error of $0.03$ means that the predicted final size is off by $3\%$ of the agents, on average over the seed fractions. Both weighted averages lower the error on average, from $0.140$ to $0.090$ with the weights $p_k$ and to $0.076$ with $q_k$, but not in every combination: both raise it in the same three, \post{gun-right} on ER $z=5$ and \post{police-defunded} on ER $z=2$ and $z=5$. The error is at least $0.029$ with $p_k$ and at least $0.042$ with $q_k$, in every combination.

\begin{table}[htbp]
\centering
\small
\begin{tabular}{llccc}
\hline
 & & plain & agent-weighted & neighbor-weighted\\
post & network & average $\fbar$ & average $\fbar_p$ & average $\fbar_q$\\
\hline
\post{migration-must-be-limited} & ER $z=2$ & $0.126$ & $0.029$ & $0.048$\\
                                 & ER $z=5$ & $0.108$ & $0.076$ & $0.073$\\
                                 & BA       & $0.190$ & $0.075$ & $0.076$\\
\post{migration-must-be-stopped} & ER $z=2$ & $0.075$ & $0.042$ & $0.044$\\
                                 & ER $z=5$ & $0.072$ & $0.046$ & $0.069$\\
                                 & BA       & $0.363$ & $0.092$ & $0.042$\\
\post{police-defunded}            & ER $z=2$ & $0.013$ & $0.045$ & $0.047$\\
                                 & ER $z=5$ & $0.068$ & $0.136$ & $0.071$\\
                                 & BA       & $0.296$ & $0.110$ & $0.109$\\
\post{gun-right}                 & ER $z=2$ & $0.182$ & $0.139$ & $0.081$\\
                                 & ER $z=5$ & $0.036$ & $0.145$ & $0.107$\\
                                 & BA       & $0.148$ & $0.148$ & $0.148$\\
\hline
mean over the twelve cases       &          & $0.140$ & $0.090$ & $0.076$\\
\hline
\end{tabular}
\caption{\textbf{No degree average removes the averaging error.} Averaging error mean$|\Delta\rho_\infty|$: the average, over the 25 seed fractions $\rho_0$ between $0.005$ and $0.30$, of the absolute difference between the final size of the cascade with the degree-resolved response $f$ and with a degree average, as on the vertical axis of Fig.~\ref{fig:error}A. All values are from the mean-field theory, for the plain average $\fbar$ used in the main text, the agent-weighted average $\fbar_p$ and the neighbor-weighted average $\fbar_q$.}
\label{tab:weights}
\end{table}

\subsection{\texorpdfstring{Cascade dynamics on Barab\'asi--Albert networks}{Cascade dynamics on Barab\'asi--Albert networks}}
\label{sec:cascade-sm}
The degree average fails most on BA (Fig.~\ref{fig:error}A). This section shows how large its error is there and how the heterogeneity of the network causes it.

On BA networks the {degree-resolved response} keeps a finite ignition threshold that the mean-field
reproduces, while the {degree-averaged response ignites the network from a vanishingly small seed} (tipping seed
$\to 0$). 
The degree-averaging error is therefore near-total below the tipping seed of the cascade with $f$ ($\Delta\rho_\infty\approx -1$, the gap between the two cascades in Fig.~\ref{fig:cascade}D), and because {$\delta F$} is $O(1)$ there it is outside the
reach of the perturbative eqs.~\eqref{eq:drho}--\eqref{eq:drho0}; the {difference of the two mean-field solutions} remains exact.

The error is this large because of how the degree average is built on a heterogeneous network. The average $\fbar$ gives every degree that occurs in the network the same weight. On BA, 147
different degrees occur, but most agents ($71\%$) have only 3 to 5 contacts: the degrees $k\ge10$ carry $94\%$ of
the weight in the average ($\fbar$), although only $11\%$ of the agents have them. Since high-degree agents need a smaller fraction $(m/k)$
of sharing contacts before they {share} (Section~\ref{sec:kernel}), the degree average ends up close to the response
of a hub. For \post{migration-must-be-limited}, for example, $\fbar$ crosses $\tfrac12$ at $x\approx0.20$, while
the agents with 3 to 5 contacts, who make up most of the network, cross it between $x=0.37$ and $x=0.46$. Applied
to every agent, the degree average is therefore far too permissive, and the {cascade with $\fbar$} ignites the network
from a vanishingly small seed{: its tipping seed is at most $0.010$, against $0.04$--$0.13$ for the degree-resolved response (Table~\ref{tab:transfer}).}

\subsection{\texorpdfstring{{A response measured in one large group}}{A response measured in one large group}}
\label{sec:transfer-sm}
Studies of well-mixed populations measure one response in a large group in which every agent sees all the others, and use it for every agent (Section~\ref{sec:cascade}). This section tests whether such a response predicts the cascade on our networks. In our measurement an agent with $K$ contacts sees $K$ others, so our largest degree, $K=155$, is the largest such group we have: an agent is told how
many of the $K$ accounts it follows have shared. On a network, this response can be given to an agent with $m$
sharing contacts out of $k$ in two ways: read at the fraction, $f_K(m/k)$, the curve measured in the group of size
$K$ evaluated at the fraction $m/k$ (interpolated linearly between the measured points); or read at the count,
$f(m,K)$. We compute these cascades with the mean-field theory of Section~\ref{sec:theory} only, without Monte Carlo simulations, because the theory reproduces them for both the degree-resolved and the degree-averaged response on every network (Fig.~\ref{fig:cascade}), and it is least accurate near a tipping point, which neither reading of the large-group response has within the range of seeds. Table~\ref{tab:transfer} gives the results.

Read at the fraction, the response ignites every network from the smallest seed, $\rho_0=0.005$: an agent with few
contacts shares as soon as one of them has shared, because one contact out of a few is a large fraction. Read at
the count, it hardly spreads beyond the seed (the final size exceeds the seed fraction by at most $0.011$, on BA,
and by less than $0.001$ on the ER networks): an agent with few contacts never reaches the count at which a member
of a large group shares. The cascade with the degree-resolved response, which the LLM population follows
(Fig.~\ref{fig:cascade}), lies between these two extremes.

The error of the {large-group} response is at least as large as that of the degree-averaged response in 23 of the 24
comparisons. Read at the count, it is larger in all twelve cases, although only marginally ($0.076$ against
$0.075$) for \post{migration-must-be-stopped} on ER $z=2$, whose cascade with $f$ does not reach one half either.
Read at the fraction, it is larger in eight cases and equal in three, on BA, where both responses ignite the
network from the smallest seed. The exception is \post{gun-right} on ER $z=2$, with $0.055$ against $0.182$. There
the cascade with $f$ also starts from a very small seed, because the measured response of this post to a single
sharer is high at low degree ($f(1,3)=0.82$, Table~\ref{tab:lowk}); read at the fraction, the large-group response
also makes an agent with few contacts share after a single sharer, so its agreement there is accidental.

The two errors differ in nature. The degree-averaged response is exact on a network in which every agent has the
same degree, and its error grows with the heterogeneity of the network (Fig.~\ref{fig:error}B). The {large-group}
response gives every agent the response of a degree that almost no agent has (on BA, $0.04\%$ of the agents have
$k\ge100$; on the ER networks, no agent has more than 18 contacts), so it fails even when all agents have the same
degree: on the regular networks from which the heterogeneity sweeps of Section~\ref{sec:hetero} start ({$k=6$, 7, 9 and 12}), its error is between {$0.19$ and $0.65$} read at the fraction, and between {$0.16$ and $0.66$} read at the count.

\begin{table}[htbp]
\centering
\small
\begin{tabular}{llcccc}
\hline
 & & tipping seed & \multicolumn{3}{c}{error mean$|\Delta\rho_\infty|$}\\
post & network & with $f$ & $\fbar$ & $k=155$, fraction & $k=155$, count\\
\hline
\post{migration-must-be-limited} & ER $z=2$ & $0.214$ & $0.126$ & $0.498$ & $0.192$\\
                                 & ER $z=5$ & $0.102$ & $0.108$ & $0.294$ & $0.548$\\
                                 & BA       & $0.061$ & $0.190$ & $0.190$ & $0.653$\\
\post{migration-must-be-stopped} & ER $z=2$ & ---     & $0.075$ & $0.601$ & $0.076$\\
                                 & ER $z=5$ & $0.211$ & $0.072$ & $0.578$ & $0.263$\\
                                 & BA       & $0.133$ & $0.363$ & $0.402$ & $0.444$\\
\post{police-defunded}            & ER $z=2$ & $0.187$ & $0.013$ & $0.449$ & $0.240$\\
                                 & ER $z=5$ & $0.135$ & $0.068$ & $0.388$ & $0.454$\\
                                 & BA       & $0.099$ & $0.296$ & $0.296$ & $0.551$\\
\post{gun-right}                 & ER $z=2$ & $0.016$ & $0.182$ & $0.055$ & $0.634$\\
                                 & ER $z=5$ & $0.058$ & $0.036$ & $0.184$ & $0.658$\\
                                 & BA       & $0.044$ & $0.148$ & $0.148$ & $0.698$\\
\hline
\end{tabular}
\caption{\textbf{A response measured in one large group mispredicts the cascade.} The response measured in the largest group ($k=155$), given to every agent and read at the fraction
$m/k$ or at the count $m$, against the degree-resolved response $f$ and the degree-averaged response $\fbar$, from
the mean-field theory. Tipping seed: the seed fraction at which the final size of the cascade with $f$ first
reaches one half (---: not reached for $\rho_0\le0.30$). Read at the fraction, the large-group response reaches one
half at the smallest seed, $\rho_0=0.005$, in every case; read at the count, it never does. Error: the average,
over the 25 seed fractions $\rho_0$ between $0.005$ and $0.30$, of the absolute difference between the final size
of the cascade with $f$ and with the approximation, as in Fig.~\ref{fig:error}A and Table~\ref{tab:weights}; the
$\fbar$ column repeats the plain-average column of Table~\ref{tab:weights}.}
\label{tab:transfer}
\end{table}

\subsection{\texorpdfstring{Choices behind the degree dependence and the heterogeneity sweeps}{Choices behind the degree dependence and the heterogeneity sweeps}}
\label{sec:choices-sm}
Fig.~\ref{fig:error} rests on two choices that the main text states briefly: the weights $q_k$ in the degree dependence of the response, and the mean degree and the seed fractions of the heterogeneity sweeps. This section shows what other weights would change and why the sweeps are set up as they are.

\emph{The weights of the degree dependence.} Eq.~\eqref{eq:dFnorm} weights each degree by $q_k$ because the propagated term of eq.~\eqref{eq:linear}, which the susceptibility $\chi$ amplifies near the tipping point, sums over degrees with these weights. With the weights $p_k$ of the direct term instead, the degree dependence orders the error less well: the rank correlation between the error and the degree dependence over the twelve post--network cases drops from $0.92$ to $0.84$, and the error of \post{gun-right} no longer grows along its line in Fig.~\ref{fig:error}A.

\begin{figure}[tbp]
  \centering
  \includegraphics[width=\textwidth]{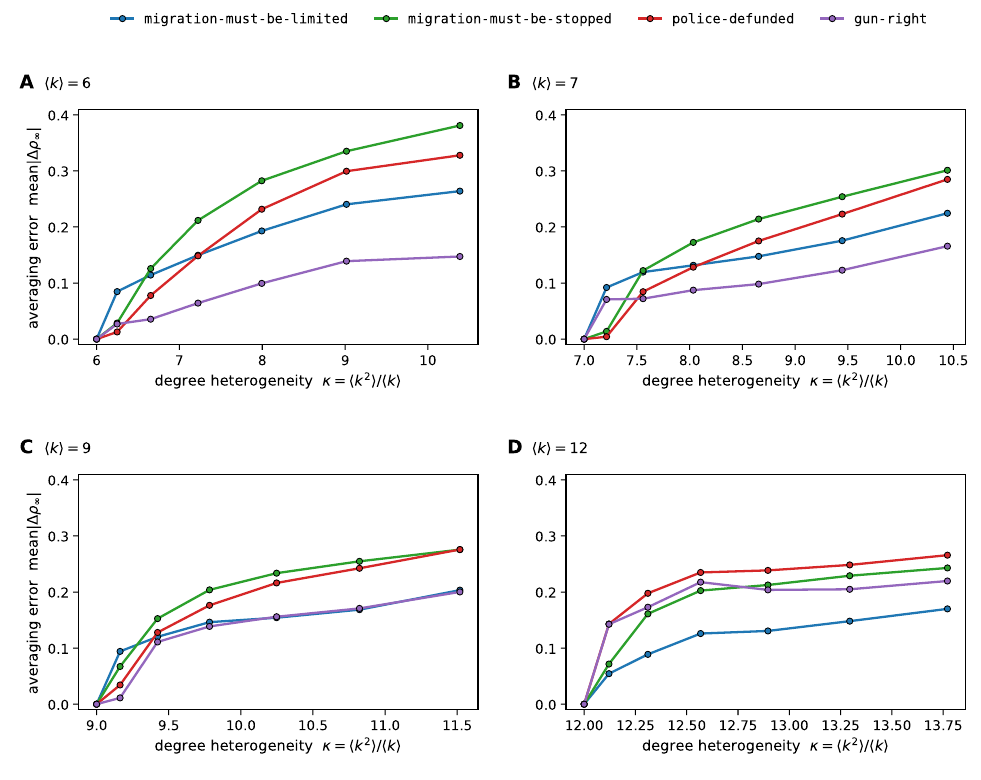}
  \caption{{\textbf{The averaging error grows with heterogeneity at every mean degree.} Averaging error mean$|\Delta\rho_\infty|$ against the degree heterogeneity $\kappa=\langle k^2\rangle/\langle k\rangle$, from the mean-field theory (601 seed fractions), on networks rewired from a regular graph with mean degree $\langle k\rangle=6$ (\textbf{A}), 7 (\textbf{B}, as in Fig.~\ref{fig:error}B), 9 (\textbf{C}) and 12 (\textbf{D}). Points: $n/N=0,0.5,\dots,3$ rewiring steps per agent. The error is zero on the regular graph, and its rank correlation with $\kappa$ is positive in all sixteen sweeps, between $0.89$ and $1.00$.}}
  \label{fig:sm_hetero}
\end{figure}

\emph{The heterogeneity sweeps.} Fig.~\ref{fig:error}B uses the mean degree $\langle k\rangle=7$, close to that of the ER $z=5$ and BA networks of Fig.~\ref{fig:error}A, so that both panels compare networks of similar density. Fig.~\ref{fig:sm_hetero} shows the sweeps at all four mean degrees. We average the error over 601 seed fractions in the window $0.005\le\rho_0\le0.30$, instead of the 25 of the Monte Carlo simulations, because on these networks the transition is discontinuous: each of the 25 seeds that falls between the two tipping seeds adds about one twenty-fifth of the jump to the average, so a 25-seed average moves in steps of about $0.036$ and can rise and fall even when the gap between the tipping seeds grows smoothly.

\subsection{\texorpdfstring{How the error depends on the mean degree}{How the error depends on the mean degree}}
\label{sec:meandegree-sm}
Section~\ref{sec:hetero} draws no conclusion about how the error depends on the mean degree, because the trend depends on how the degree average is built. This section shows the trend for the plain and the two weighted averages (Section~\ref{sec:weights-sm}), and that the growth of the error with heterogeneity survives this choice.

\emph{The trend with mean degree depends on the average.} In the heterogeneity sweeps of Fig.~\ref{fig:sm_hetero}, at the same heterogeneity, $c_v^2=\langle k^2\rangle/\langle k\rangle^2-1=0.1$, and averaged over the four posts, the error grows with the mean degree for the plain average, from $0.082$ at $\langle k\rangle=6$ to $0.205$ at $\langle k\rangle=12$, but falls for the neighbor-weighted one, from $0.056$ to $0.028$, and for the agent-weighted one, from $0.069$ to $0.018$. The growth of the error with heterogeneity survives the choice: its rank correlation with $\kappa$ is positive in fifteen of the sixteen sweeps with the neighbor-weighted average, and in the exception, a sweep of \post{migration-must-be-stopped}, the error stays flat or falls slightly; with the agent-weighted average it is positive in eleven of the sixteen, all except the four sweeps of \post{migration-must-be-stopped} and the sweep of \post{migration-must-be-limited} at $\langle k\rangle=12$. 

\subsection{\texorpdfstring{{Expansion of the averaging error}}{Expansion of the averaging error}}
\label{sec:methods-error}\label{sec:derivation-sm}
Section~\ref{sec:closedform} explains the errors of the degree average with a first-order expansion, eq.~\eqref{eq:linear}, and tests factorization with a single number $g$ for the degree dependence of the response. This section derives the expansion and the shift of the tipping seed, shows why no single number can order the errors, and defines $g$.

\emph{The error to first order.} In the mean-field theory, eq.~\eqref{eq:fixedpoint}, the probability $q$ that a link leads to an active agent solves $q=G[F](q)$, and the final size is $\rho_\infty=R[F](q^{*})$, with $q^{*}$ the smallest solution. The functions $G$ and $R$ are sums over degrees of the binomial averages $A$ and $H$ of the response, eqs.~\eqref{eq:Akq} and~\eqref{eq:H}, so they are linear in the response. Hence $G[F]=G[\bar F]+\delta G$ and $R[F]=R[\bar F]+\delta R$, with $\delta G(q)=(1-\rho_0)\sum_k q_k\,\delta A(k,q)$ and $\delta R(q)=(1-\rho_0)\sum_k p_k\,\delta H(k,q)$, where $\delta A=A[F]-A[\bar F]$ and $\delta H=H[F]-H[\bar F]$. At the same $q$, $\delta G$ is the change in the probability that a link leads to an active agent, and $\delta R$ the change in the probability that an agent is active. Let $q_0(\rho_0)$ solve the fixed point $q_0=G[\bar F](q_0)$ with $\fbar$. To first order in $\delta F$, the fixed point with $f$ is $q^{*}=q_0+\Delta q$. Inserting it into $q=G[\bar F](q)+\delta G(q)$ and keeping only first-order terms, $q_0$ cancels against $G[\bar F](q_0)$, and what remains is $\Delta q=G'[\bar F](q_0)\,\Delta q+\delta G(q_0)$, where $G'$ is the derivative of $G$ with respect to $q$. Solving for $\Delta q$ gives
\begin{equation}
  \Delta q \;=\; \chi(\rho_0)\,\delta G(q_0), \qquad \chi(\rho_0)\equiv\frac{1}{1-G'[\bar F](q_0)}, \qquad G'[\bar F](q_0)=(1-\rho_0)\sum_k q_k\,\partial_q A[\bar F](k,q_0).
  \label{eq:dq}
\end{equation}
The susceptibility $\chi$ says how strongly the fixed point responds to a change in $G$: a link that leads to an extra active agent makes, on average, $G'$ further links lead to active agents, these make $G'^2$ more, and so on, and $\chi=1+G'+G'^2+\dots$ sums this chain. It diverges at the tipping point, where $G'[\bar F](q_0)\to1$. Inserting $q^{*}$ into $R[F]$ and keeping again the first-order terms gives $R[F](q^{*})-R[\bar F](q_0)=\delta R(q_0)+R'[\bar F](q_0)\,\Delta q$, where $R'$ is the derivative of $R$ with respect to $q$. The error in the final size is therefore
\begin{equation}
  \Delta\rho_\infty \;=\; \delta R(q_0) \;+\; R'[\bar F](q_0)\,\chi(\rho_0)\,\delta G(q_0) .
  \label{eq:drho}
\end{equation}
The first term is the direct change: at the same probability $q_0$ that its neighbors are active, an agent with the response $F$ is active with a different probability than one with $\bar F$. The second term is the propagated change: the deviation of the response also changes the probability that a link leads to an active agent, which moves the fixed point by $\Delta q$; the susceptibility amplifies this change, and $R'$ turns it into a change of the final size. Near the tipping point, where $\chi$ is large, the propagated term dominates. Writing $\delta R$ and $\delta G$ as sums over $\delta F$ with eqs.~\eqref{eq:Akq} and~\eqref{eq:H} gives eq.~\eqref{eq:linear} of the main text. At small seed fractions, the sign of $\delta G(q_0)$ computed from the measured response matches the sign of the error in all twelve post--network combinations: $\delta G$ is negative, and the degree average overpredicts the cascade, except for \post{gun-right} on ER $z=2$.

\emph{The shift of the tipping seed.} Where the cascade jumps, the final size is discontinuous at the tipping seed, and between the tipping seeds of the two cascades their final sizes differ by an amount that a first-order expansion cannot capture. There the useful observable is the shift of the tipping seed, $\Delta\rho_0^{*}=\rho_0^{*}[f]-\rho_0^{*}[\fbar]$. The tipping seed of the theory is located by two conditions, $q=G[\bar F](q)$ and $G'[\bar F](q)=1$: the curve $G[\bar F](q)$ touches the diagonal, so that the smallest fixed point disappears and the final size jumps (a saddle-node bifurcation). With $f$ instead of $\fbar$, the curve moves by $\delta G$, and the seed fraction at which it touches the diagonal moves by $\Delta\rho_0^{*}$. Because the curve is tangent to the diagonal at the touching point $q_c$, a small change of $q$ leaves the first condition unchanged to first order, so the change of the seed alone must compensate $\delta G$: $\partial_{\rho_0}G[\bar F](q_c)\,\Delta\rho_0^{*}+\delta G(q_c)=0$, which gives
\begin{equation}
  \Delta\rho_0^{*} \;=\; -\,\frac{\delta G(q_c)}{\partial_{\rho_0}G[\bar F](q_c)},
  \qquad \partial_{\rho_0}G[\bar F](q_c)=1-\sum_k q_k A[\bar F](k,q_c)>0 ,
  \label{eq:drho0}
\end{equation}
with $q_c$ the fixed point with $\fbar$ at its tipping seed. The derivative is positive because a seed is active for certain, while any other agent is active only with probability $A\le1$, so a larger seed raises $G$. In all our discontinuous cases the final size jumps at this seed from below one half to above it, so it is the tipping seed defined in Section~\ref{sec:observed}. Written out, it is again a weighted sum of $\delta F$, $\Delta\rho_0^{*}=\sum_{m,k}W'(m,k)\,\delta F(m,k)$ with $W'(m,k)=-(1-\rho_0)q_k b_m(k{-}1,q_c)/\partial_{\rho_0}G[\bar F](q_c)$. When $\delta G(q_c)<0$, as in all our discontinuous cases, the tipping seed with $f$ is larger than the one with $\fbar$: the degree average underestimates how large a seed is needed to ignite the network.

\emph{Why no single number orders the errors.} The sum of eq.~\eqref{eq:linear} is a scalar product, $\Delta\rho_\infty=\langle W,\delta F\rangle$, with $\langle a,b\rangle=\sum_{m,k}a(m,k)\,b(m,k)$. With the norm $\|a\|=\langle a,a\rangle^{1/2}$, the error can be written as $\Delta\rho_\infty=\|W\|\,\|\delta F\|\cos\theta$, where $\theta$ is the angle between $W$ and $\delta F$. This splits the error into three factors. The first, $\|\delta F\|$, is the size of the response deviation. The second, $\|W\|$, is the amplification by the network: it contains $p_k$, $q_k$ and the susceptibility $\chi$, and it diverges at the tipping point. The third, $\cos\theta$, is the alignment between $\delta F$ and $W$: it is large when the response deviates where the cascade samples it. A norm of $\delta F$ sees only the first factor. The second depends on the network, not on the deviation of the response, and the third depends on the shape of $\delta F$, which changes from post to post. Therefore no single number that summarizes $\delta F$ can order the errors of different posts, which is the failure of factorization of Section~\ref{sec:factorization}. By the Cauchy--Schwarz inequality, $|\Delta\rho_\infty|\le\|W\|\,\|\delta F\|$, and the bound is reached only when $\delta F$ is aligned with $W$.

\emph{The single number $g$.} For the factorization test of Section~\ref{sec:factorization} we summarize the degree dependence of the response by a single number $g$, the root-mean-square departure of the measured response from a
pure fraction rule: $g^2$ is the average, over all measured $(m,k)$, of $[f(m,k)-\bar f_{\rm all}(m/k)]^2$, where
$\bar f_{\rm all}$ is the average of the response curves over all measured degrees. We choose this $g$ because it
depends only on the {post} and not on the network, which is what a factorization requires. On each network of the heterogeneity sweeps where the error exceeds $0.01$, we divide the largest value of error$/g$ among the four posts by the smallest; the median of this quotient over these networks is $2.1$, the factor of about two of Section~\ref{sec:factorization}. {A counterexample to factorization is} \post{migration-must-be-stopped}: from $n/N=1.5$ on it has the largest error of the four {posts} at
$\langle k\rangle=7$ (Fig.~\ref{fig:error}B), but a smaller $g$ than \post{migration-must-be-limited}
($0.131$ against $0.147$).

\subsection{\texorpdfstring{Validation of the first-order expansion of the error against the Monte Carlo simulations}{Validation of the first-order expansion of the error against the Monte Carlo simulations}}
\label{sec:eq7-table}\label{sec:newtheory-sm}
Section~\ref{sec:closedform} states that the first-order expansion reproduces the measured error where the two cascades stay close. This section gives the evidence.

\begin{figure}[htbp]
  \centering
  \includegraphics[width=\textwidth]{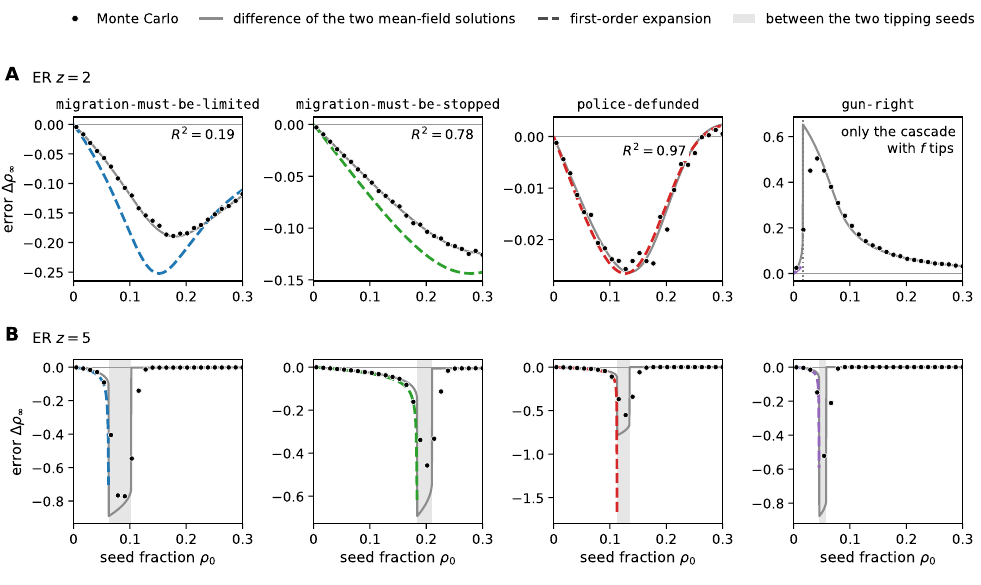}
  \caption{{\textbf{The first-order expansion against the Monte Carlo simulation, post by post.} Error $\Delta\rho_\infty$ of the degree-averaged response against the seed fraction $\rho_0$, for the four posts (columns). Black dots: Monte Carlo simulations; gray line: difference of the two mean-field solutions; dashed lines, colored by post as in Fig.~\ref{fig:error}: first-order expansion, eq.~\eqref{eq:drho}, computed from the solution with $\fbar$ and the measured response alone. (\textbf{A}) ER $z=2$, continuous transition, with the $R^2$ of the expansion against the Monte Carlo simulations from the smallest seed to the extremum of the error. For \post{gun-right} only the cascade with $f$ tips (dotted line), so the error is $O(1)$ beyond the tipping seed and the expansion is shown only below it. (\textbf{B}) ER $z=5$, discontinuous transition; gray band: the window between the two tipping seeds, whose shift Table~\ref{tab:eq7} lists.}}
  \label{fig:newtheory}
\end{figure}

{Figure~\ref{fig:newtheory} compares the expansion with the Monte Carlo simulation for each post: the pointwise error, eq.~\eqref{eq:drho}, on ER $z=2$, where the transition is continuous (panel A), and the tipping-seed shift, eq.~\eqref{eq:drho0}, on ER $z=5$, where the transition is discontinuous and the shift is the well-posed observable (panel B).} 
{On BA networks the two cascades do not stay close, because $\delta F$ is of order one, and the expansion does not apply (Section~\ref{sec:cascade-sm}).}

\emph{Continuous transition (ER $z=2$).} {In Fig.~\ref{fig:newtheory}A, eq.~\eqref{eq:linear}} is compared with the Monte Carlo error for the three posts for which both responses stay continuous, \post{migration-must-be-limited}, \post{migration-must-be-stopped} and \post{police-defunded}. We compute $R^2$ from the
smallest seed up to the extremum of the measured error, because this range is where the {expansion}
is expected to hold; beyond the extremum the susceptibility $\chi$ of eq.~\eqref{eq:linear}, which amplifies the response deviation near the tipping point, grows and the leading-order expansion degrades. It
follows the error through and past its minimum for \post{police-defunded} ($R^2=0.97$). For {\post{migration-must-be-stopped}}
it reproduces the shape of the error but overestimates its size, by about $14\%$ at the deepest point ($R^2=0.78$), and for \post{migration-must-be-limited} by about $34\%$ ($R^2=0.19$). The overestimate grows with the depth of the error, which is $0.03$ for \post{police-defunded}, $0.13$ for \post{migration-must-be-stopped} and $0.19$ for \post{migration-must-be-limited}, as one expects from an expansion that keeps only the first order in $\delta F$, while the difference of the two mean-field solutions follows the Monte Carlo error for all three. {For \post{gun-right} the error is $O(1)$ beyond the tipping seed of the cascade with $f$, where the expansion does not apply; 
once the cascade with $\fbar$ has also spread, at larger seeds, the difference of the two mean-field solutions tracks the Monte Carlo error closely.}

\emph{Discontinuous transition (ER $z=5$).} Table~\ref{tab:eq7} lists the {tipping-seed shift} $\Delta\rho_0^{*}$ predicted by the closed form
eq.~\eqref{eq:drho0} against the Monte Carlo shift. The prediction is accurate for all four {posts}; it falls $\sim\!9\%$ short
only for \post{migration-must-be-limited}, the {post} with the largest shift (leading-order truncation). The theory and the Monte Carlo simulations use the same tipping seed, the seed fraction at which the final size crosses one half (Section~\ref{sec:observed}). The theory locates it where its smallest fixed point disappears and the final size jumps through one half (Section~\ref{sec:methods-error}); the Monte Carlo simulations locate it where the average final size crosses one half.

\begin{table}[htbp]
\centering
\small
\begin{tabular}{lccc}
\hline
{post} (ER $z=5$) & {expansion, eq.~\eqref{eq:drho0}} & Monte Carlo & {mean-field difference, all orders}\\
\hline
\post{migration-must-be-stopped} & $+0.027$ & $+0.026$ & $+0.026$\\
\post{police-defunded}            & $+0.023$ & $+0.023$ & $+0.022$\\
\post{gun-right}                 & $+0.013$ & $+0.013$ & $+0.013$\\
\post{migration-must-be-limited} & $+0.035$ & $+0.039$ & $+0.040$\\
\hline
\end{tabular}
\caption{{\textbf{The expansion predicts the shift of the tipping seed.} Shift $\Delta\rho_0^{*}$ of the tipping seed on ER $z=5$, from the first-order expansion, eq.~\eqref{eq:drho0}; from the Monte Carlo simulations; and from the difference of the two mean-field solutions, to all orders in $\delta F$. The expansion matches the Monte Carlo simulations for every post, within $10\%$.}}
\label{tab:eq7}
\end{table}

\end{document}